\documentclass[acmsmall,nonacm]{acmart}
\usepackage{xcolor}
\usepackage{colortbl, boldline}
\usepackage{amsthm}
\usepackage{booktabs}
\usepackage{caption}
\usepackage{adjustbox}
\usepackage{array}
\usepackage{lineno}
\usepackage{graphicx}
\usepackage{capt-of}    
\usepackage{tabularx}
\usepackage{subfig}
\usepackage{listings}
\usepackage{longtable}
\usepackage{bbding}
\usepackage{fancybox}
\usepackage[xcolor]{changebar}
\usepackage{parcolumns}

\usepackage{pifont}

\newcolumntype{R}[2]{%
    >{\adjustbox{angle=#1,lap=\width-(#2)}\bgroup}%
    l%
    <{\egroup}%
}

\newcommand{\changed}[1]{%
    #1
  }%

\AtBeginDocument{%
  \providecommand\BibTeX{{%
    \normalfont B\kern-0.5em{\scshape i\kern-0.25em b}\kern-0.8em\TeX}}}

\setcopyright{acmcopyright}
\copyrightyear{2018}
\acmYear{2022}
\acmDOI{10.1145/1122445.1122456}

\begin{document}

\title{Algorithm Selection for Software Verification using Graph Neural Networks}

\author{Will Leeson}
\orcid{0000-0003-2403-9295}
\author{Matthew B. Dwyer}
\affiliation{%
  \institution{University of Virginia}
  \streetaddress{85 Engineer's Way}
  \city{Charlottesville}
  \state{Virginia}
  \country{USA}
  \postcode{22903}
}

\renewcommand{\shortauthors}{Leeson and Dwyer}
\begin{abstract}

The field of software verification has produced a wide array of algorithmic techniques that can prove a variety of properties of a given program. It has been demonstrated that the performance of these techniques can vary up to 4 orders of magnitude on the same verification problem. Even for verification experts, it is difficult to decide which tool will perform best on a given problem. For general users, deciding the best tool for their verification problem is effectively impossible.

In this work, we present \textsc{Graves}, a selection strategy based on graph neural networks (GNNs). \textsc{Graves} generates a graph representation of a program from which a GNN predicts a score for a verifier that indicates its performance on the program.

We evaluate \textsc{Graves} on a set of 10 verification tools and over 8000 verification problems and find that it improves the state-of-the-art in verification algorithm selection by 12\%, or 8 percentage points. Further, it is able to verify 9\% more problems than any existing verifier on our test set. 
Through a qualitative study on model interpretability, we find strong evidence that the \textsc{Graves}' model learns to base its predictions on factors that relate to the unique features of the algorithmic techniques.
\end{abstract}


\maketitle

\section{Introduction} \label{intro}

Given a program, $P$, and a correctness specification, $\phi$, formal verification seeks to determine whether the executable program behavior is consistent with the specification, $P\models\phi$. 
In practice, $\phi$ can take many forms, such as that all assertions hold, that memory is used safely, or a guarantee that a program will terminate. 
Verification tools must prove that all feasible program executions do not violate $\phi$. 
\changed{
Many verification techniques have been introduced~\cite{biere2009bounded,clarke2000counterexample,de2003bounded,flanagan2002predicate,beyer2008program,chase1990analysis,henzinger2002lazy}, with different tools implementing different subsets of techniques~\cite{beyer2022progress}, leading to a diversity in the performance of verifiers. In recent competition settings, 19 of the 20 competing tools built for verifying C programs were able to solve somewhere between 4 and 500 verification problems that no other tool could solve~\cite{beyer2021software}.
Of the 15000 verification problems in the competition, nearly 10\% could be solved by only one of the competing verifiers. 

In order to use these verifiers, developers must learn how to write specifications the tools can consume. If each verifier uses a different specification language, then developers would have to learn each language to make use of the verification field's diversity. Generally, this results in developers learning a single specification language and only making use of a few, or even one, verifier(s). In recent years, there has been an effort to standardize the language in which $\phi$ is specified for verification tools~\cite{baudin2021acsl,beyer2022static,leavens2006preliminary}.
This shifts the question from ``Which specification language should I learn to use?'' to ``Which verifier should I use for this program?''.

Deciding which tool is best suited to verify a specific piece of software can be difficult for an expert in the field of formal software verification, let alone a non-expert software developer. 
To decide which verifier is most appropriate, the user of the verification tool must have knowledge of the possible execution behaviors of the program. Oftentimes, software is built by teams of developers, making it difficult to be aware of the full behavior of the system. Further, verification may be carried out by a separate team~\cite{feldt2010challenges}, which would not be privy to the intricate details of the implementation.  Moreover, the user must also be aware of what techniques each verifier implements, the behaviors where each technique excels, the behaviors where each technique struggles, and what behavior causes the verifier to trigger the different techniques the tool implements. Requiring users of verifiers to be experts in the field of verification tools is impractical.}

\begin{figure}
\noindent\begin{minipage}{.45\textwidth}
\begin{lstlisting}[caption=Program A,frame=tlrb,language=C]{Name}
1.  int x = 0;
2.  int y = 1;
3.  while(x==0){
4.      x = x + 2;
5.  }
6.  y++;
7.  assert(x==y);
\end{lstlisting}
\end{minipage}\hfill
\begin{minipage}{.45\textwidth}
\begin{lstlisting}[caption=Program B,frame=tlrb,language=C]{Name}
1.  int x = 0;
2.  int y = 1;
3.  while(x==0){
4.      y++;
5.  }
6.  x = x + 2;
7.  assert(x==y);
\end{lstlisting}
\end{minipage}
\caption{The above programs are composed of the same statements, but the order of lines 4 and 6 are altered. Program A is easily verified as true, while some tools will struggle to verify Program B as they need to prove that the while loop will never terminate. Nonetheless, feature vector based algorithm selectors would produce identical feature vectors for both of these programs.}
\label{fig:feature_vector}
\end{figure}

\changed{
To relieve developers of this burden, algorithm selectors for program verification have been introduced~\cite{richter2020algorithm,richter2020attend,demyanova2017empirical,tulsian2014mux}. Given a program and a specification, an algorithm selector will predict from a suite of verifiers which verifier is mostly likely to most efficiently check the specification. Most algorithm selectors employ a machine learning model to make predictions, although there has been some work using manually crafted rules to select verification techniques inside individual tools~\cite{darke2021veriabs,beyer2018strategy}. To train a model, the designer of the selector needs to represent the program in a way amenable to using machine learning.  

The earliest verifier selectors represented programs using ``feature vectors''. A feature vector is a vector of statistics, called features, which an expert has deemed important to differentiating between the performance of verifiers. Common features include the number of loops in the program, the number of inputs to the program, or the number of pointers in the program~\cite{demyanova2017empirical,tulsian2014mux}. An issue with feature vector-based approaches is that they do not account for the interaction between elements of the program. Figure~\ref{fig:feature_vector} shows two programs, A and B, which are nearly identical with the exception that lines 4 and 6 are swapped. Since they contain the same count of features, they will have identical feature vectors. Program A can easily be verified. The while loop will execute once, incrementing x by 2. Line 6 will increment y by 1, and the assertion on line 7 will be satisfied. Program B can also be verified to show the assertion will not be violated, but for a very different reason. The while loop will never end, meaning the assertion statement will never be reached. To prove this, a verifier must prove that the loop condition will never be false. Some verifiers, like bounded model checkers, would easily be able to verify Program A, but would not be able to verify Program B as they can only provide assurances up to a finite number of iterations. However, a feature vector based approach would produce the same prediction for these programs, since their feature vectors are the same.

\begin{figure}
    \includegraphics[width=\linewidth]{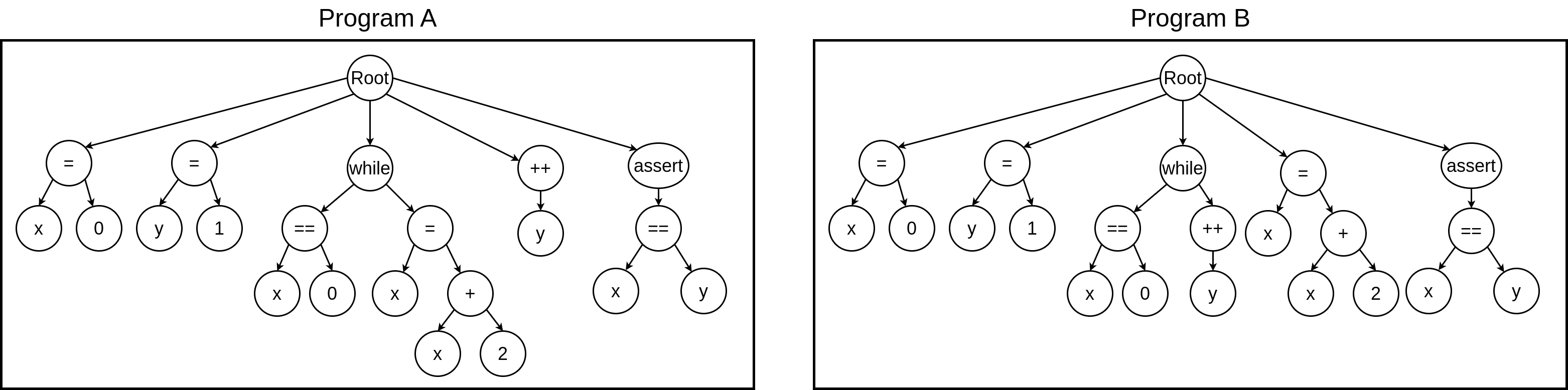}
\caption{The Abstract Syntax Trees (AST) of the programs in Figure~\ref{fig:feature_vector}. State-of-the-art selectors base their representations of programs on the AST. Programs A and B would have identical feature vectors, but their ASTs differ, allowing the graph-based selectors to differentiate between the two.}
\label{fig:Program-ASTs}
\end{figure}

To address this issue, more advanced selectors use graph representations of the program based on the program's abstract syntax tree (AST)~\cite{richter2020attend,richter2020algorithm}. Figure~\ref{fig:Program-ASTs} shows the ASTs for both Program A and Program B. The small differences between programs A and B are reflected in the AST, as would any other syntactic change. Since the graphs differ, graph based selectors will have a chance to differentiate between these two programs, allowing them to learn to select different verifiers for them.  

CST~\cite{richter2020attend} is a graph-based selector which uses a program's AST to decide which verifier to select. It uses an attention mechanism~\cite{vaswani2017attention} to learn which components of the AST are related to verifier performance. Since CST only uses the AST to represent programs, it is not \emph{expressive} enough to directly capture the data or control dependencies which are vital to deciding if a property is satisfied. In order to prove that the assertion in Program B is not violated, a verifier must identify that x's value will never change in the body of the loop. Control dependence identifies that only the statement on line 4 will be executed in the body of the while loop. Data dependence shows that only lines 1 and 6 alter the value of x. With this information, the verifier can determine x's value will not be altered in the loop. Meaning, the loop will never end, and the assertion will never be evaluated as false. Since CST does not incorporate this data, it will struggle to identify which verifier is best equipped to handle the given program. It is important that a graph representation is \emph{expressive} enough to identify relationships between nodes that influence the verification condition.

WLJ~\cite{richter2020algorithm} is another graph-based approach based on the program's AST, but it adds control flow, call, and data dependence edges to the nodes of the AST. Unlike CST, WLJ directly incorporates information verifiers use to check properties, making WLJ better suited to differentiate between verifier performance. Over a series of iterations, for each node n in a graph G, WLJ concatenates the values of n's neighboring nodes to n's value. Using G with the new node values, it attempts to look for similarities between G and the graphs it was trained on to decide which verifier to select. WLJ's main limitation is that it is not \emph{adaptive}. When WLJ performs the concatenation step, it treats all nodes equally. Further, it looks for similarities amongst all subgraphs of a user defined size. Not all statements affect a program's verification cost equally. For example, in Program B, line 6 has no influence towards determining if the assertion is violated since it will never be executed, while the loop's body is crucial, as it shows x is never altered and the loop condition is never falsified. Under WLJ's approach, these lines will have the same importance during both the concatenation step and the subgraph comparisons. It is important that a selector is able to \emph{adapt} to a given problem and identify the components of a program which are most influential to proving the property holds.

In this work, we introduce \textsc{Graves}, an \emph{expressive} and \emph{adaptive} approach to algorithm selection for program verification. Given a program $P$ and a specification $\phi$, \textsc{Graves} ranks a portfolio of verification tools based on their ability to accurately and efficiently verify $P$. Through an automated process, $P$ is converted into a graph, $G$, which is constructed from the program's AST with control flow, data flow, and function call edges added. A graph neural network encodes $G$ into a single vector which a neural network uses to produce a fitness score for each verifier in \textsc{Graves}' portfolio.

\textsc{Graves} encodes programs using a graph that directly incorporates information verifiers use to perform their analysis, allowing it to be more \emph{expressive} than CST which uses only the AST. \textsc{Graves} uses state-of-the-art graph neural networks (GNN) layers that learn to \emph{adapt} \textsc{Graves}' graph encoding to the given program. Like WLJ, \textsc{Graves} alters the value of each node in the graph using its neighbors. Unlike WLJ, it learns to weigh each node, effectively boosting the values of nodes it learns to be important. Using a pooling layer, \textsc{Graves} then combines the nodes of the graph together into a single, fixed sized vector. This pooling layer makes use of an attention mechanism, allowing \textsc{Graves} a chance to identify components of the program graph that are impactful in deciding verifier performance.

We prototyped \textsc{Graves} inside the verification framework CPAChecker~\cite{beyer2011cpachecker} and entered it into SV-Comp 2021 under the name \textsc{Graves-CPA}~\cite{GRAVES-SVCOMP22}\footnote{https://github.com/will-leeson/cpachecker}. In competition, it beat PeSCo, WLJ's CPAChecker-based implementation, in the ``Software Systems'' category, which evaluates tools on large, realistic software projects. \textsc{Graves-CPA} used a less sophisticated CPA configuration than PeSCo, with \textsc{Graves-CPA} utilizing a smaller portfolio of verification techniques than PeSCo. \textsc{Graves-CPA} also allowed techniques predicted to fail to run instead of exiting early. In spite of these limitations, the success of \textsc{Graves-CPA} motivated us to build on these ideas by incorporating more efficient and effective GNN layers into \textsc{Graves}. 

The contributions of this paper are the following:
\begin{itemize}
    \item In Section~\ref{approach}, we introduce \textsc{Graves}, an approach to verifier selection a using GNN designed with state-of-the-art graph neural network layers;
    \item In Section~\ref{selectionRelated}, we provide a discussion of existing techniques to verifier selection, their limitations, and how \textsc{Graves} avoids said limitations;
    \item In Section~\ref{RQ1}, we discuss the limitations of existing metrics for evaluating verifier selectors, and suggest a metric which evaluates their performance in a way that is more meaningful to end users;
    \item In Section~\ref{RQ3}, we perform an ablation study on \textsc{Graves}' GNN and find each component adds to \textsc{Graves}' performance; and
    \item In Section~\ref{model_interpretability}, we perform a first-of-its-kind qualitative study into the interpretability of \textsc{Graves}' GNN which identifies portions of programs that an expert would use to select a verification technique and shows that each component of \textsc{Graves}' graph representation adds value to \textsc{Graves'} predictions
\end{itemize}
}

\section{Background} \label{background}
In this section, we present background information on both automated software verification and graph neural networks.

\subsection{Automated Software Verification Tools} \label{verificationBackground}
Developments in the field of automated software verification have led to a diverse set of verification techniques. 
Each technique has strengths and weaknesses. 
Many model checker based tools convert programs into an SMT formula in an attempt to prove that no values of the free, or input, variables lead to a property violation. 
Thus, the power of the tool hinges on the SMT solver's ability to solve the complex formula they provide it. Abstract interpreters use abstract domains to characterize the variables and paths in programs. 
If an abstract domain is not precise enough to capture the semantics of the program, it may over-approximate the program's behavior and produce a false-positive, showing the program is not safe when it is.

Most modern tools do not implement a single verification technique. 
Instead, they combine techniques to make a more sophisticated verifier. 
For example, the CPAChecker framework allows developers to build their own verification tool by combining pre-implemented techniques using a configuration file~\cite{beyer2011cpachecker}. 
Because tools implement different sets of techniques, there is an algorithmic diversity which allows some verifiers to excel where others fail.

\begin{table}[]
\caption{Techniques implemented by several tools for the 2018 SV-Comp}
\resizebox{\textwidth}{!}{
\begin{tabular}{@{}l
>{\columncolor[HTML]{EFEFEF}}c c
>{\columncolor[HTML]{EFEFEF}}c c
>{\columncolor[HTML]{EFEFEF}}c c
>{\columncolor[HTML]{EFEFEF}}c c
>{\columncolor[HTML]{EFEFEF}}c c@{}}
\toprule
\textbf{\begin{tabular}[l]{@{}l@{}}Verification \\ Tool\end{tabular}} & {\color[HTML]{333333} \textbf{CEGAR}} & \textbf{\begin{tabular}[c]{@{}c@{}}Predicate \\ Abstraction\end{tabular}} & \textbf{\begin{tabular}[c]{@{}c@{}}Symbolic \\ Execution\end{tabular}} & \textbf{\begin{tabular}[c]{@{}c@{}}Bounded \\ Model \\ Checking\end{tabular}} & \textbf{k-Induction} & \textbf{\begin{tabular}[c]{@{}c@{}}Interval \\ Analysis\end{tabular}} & \textbf{\begin{tabular}[c]{@{}c@{}}Lazy \\ Abstraction\end{tabular}} & \textbf{\begin{tabular}[c]{@{}c@{}}Interpolation\end{tabular}} & \textbf{\begin{tabular}[c]{@{}c@{}}Automata \\ Based \\ Analysis\end{tabular}} & \textbf{\begin{tabular}[c]{@{}c@{}}Ranking \\ Function\end{tabular} }\\ \arrayrulecolor{black}\hline
2LS & {\color[HTML]{333333} } &  &  & \CheckmarkBold & \CheckmarkBold & \CheckmarkBold &  &  &  & \CheckmarkBold \\ \arrayrulecolor{lightgray}\hline
CBMC & {\color[HTML]{333333} } &  &  & \CheckmarkBold &  &  &  &  &  &  \\ \hline
CPA-Seq & {\color[HTML]{333333} \CheckmarkBold} & \CheckmarkBold &  &  &  &  & \CheckmarkBold & \CheckmarkBold &  & \textbf{\CheckmarkBold} \\ \hline
DepthK & {\color[HTML]{333333} } &  &  & \CheckmarkBold & \CheckmarkBold &  &  &  &  &  \\ \hline
ESBMC-Kind & {\color[HTML]{333333} } &  &  & \CheckmarkBold & \CheckmarkBold &  &  &  &  &  \\ \hline
ESBMC-Incr & {\color[HTML]{333333} } &  &  & \CheckmarkBold &  &  &  &  &  &  \\ \hline
Symbiotic & {\color[HTML]{333333} } &  & \CheckmarkBold &  &  & \CheckmarkBold &  &  &  &  \\ \hline
U. Automizer & {\color[HTML]{333333} \CheckmarkBold} & \CheckmarkBold &  &  &  &  & \CheckmarkBold & \CheckmarkBold & \CheckmarkBold &  \\ \hline
U. Kojak & {\color[HTML]{333333} \CheckmarkBold} & \CheckmarkBold &  &  &  &  & \CheckmarkBold & \CheckmarkBold &  &  \\ \hline
U. Taipan & {\color[HTML]{333333} \CheckmarkBold} & \CheckmarkBold &  &  &  &  & \CheckmarkBold & \CheckmarkBold & \CheckmarkBold &  \\ \arrayrulecolor{black}\hline
\end{tabular}}
\label{table:diversity}
\end{table}

The Competition on Software Verification (SV-Comp) is an annual event that evaluates verification tools on a diverse set of benchmarks, covering many program behaviors and several verification properties. 
In the most recent competition, SV-Comp 2021, 
48 different verification tools competed, using some subset of over 20 techniques~\cite{beyer2022progress}.
Table~\ref{table:diversity} provides an abbreviated look at the diversity of algorithmic implementations of tools at SV-Comp 2018, which we use in our study. 
While there are ``winners'' of the overall competition and the different categories, there is no single verifier that does best on all programs. 
This has motivated the creation of algorithm selectors using suites of tools from the competition. 
In fact, there are tools which compete using algorithm selectors ~\cite{richter2019pesco, darke2021veriabs,GRAVES-SVCOMP22}.

\subsection{Machine Learning} \label{MLBackground}
Machine learning is used to solve tasks that would be effectively impossible to program. For example, autonomous driving cars would require a complex series of conditionals to handle any given scenario a driver may encounter on the road. Instead of codifying a system of rules an autonomous car should follow when driving, machine learning has been used to learn how the car should react to a given scenario~\cite{sallab2017deep,kuutti2020survey}. The power of machine learning techniques are their ability to learn complex patterns in large corpora of data to make accurate predictions.

A simple, yet effective, machine learning technique for classification is the support vector machine (SVM)~\cite{boser1992training}. 
SVMs use training data to learn a boundary which maximizes the margin between the boundary and the data points of the two classes. 
When a new data point is presented to the SVM, it is classified based on which side of the boundary it lies. 
SVMs can be generalized to multiclass classification problems as well.

One of the core concepts in machine learning is the idea of an artificial neural network (ANN) ~\cite{mcculloch1943logical}. 
These networks are a series of layers of nodes with connections between layers. 
Data is input into the network and flows through the network.
As the data passes through the layers, calculations are performed on the data until the final layer is reached. 
The output is then used to answer the task the network is meant to solve. 
These networks go through a training phase where the calculations the networks perform are iteratively tuned, typically through a process called backpropagation~\cite{rumelhart1986learning}.

Traditional machine learning techniques leverage the fact that the data they operate on is of a consistent size. 
For example, bit-mapped image encodings have a consistent  dimension and ordering. 
The networks learn to make calculations accordingly. 
Recurrent neural networks (RNNs) allow for variable sized input, typically streams of data, but they still leverage the fact that data has a set ordering or pattern. 
RNNs maintain a state which is updated as data is input to it.
Graph data, in general, has no set ordering or size which makes it problematic for SVMs, ANNs and RNNs.

\subsubsection*{Graph Neural Networks}\label{GNN_Background}

Introduced in Scarselli et al.~\cite{scarselli2008graph}, graph neural networks aim not only to capture the information in the nodes of the graph, but also the connections, or edges, between them. 
An interesting observation in the foundational work is that GNNs can be thought of as a generalization of RNNs. In an RNN, data flows in a linear fashion. 
As data is fed through, calculations are made and the state is updated. 
With GNNs, this must be augmented.
Each node in the graph maintains a state vector. To encode the structure of the graph, the state of each node is passed along the edges of the graph through a process called ``message passing''. There are a variety of message passing layers which take inspiration from convolutional neural networks (CNNs)~\cite{kipf2016semi,thekumparampil2018attention,du2017topology}, RNNs~\cite{li2015gated,velivckovic2017graph,brody2021attentive}, and core concepts in graph theory~\cite{weisfeiler1968reduction,morris2019weisfeiler}. 
Initially, message passing was performed until a fixed point was reached. This required assumptions be made about the graph and the message passing layer. In modern GNNs, these restrictions are lifted, and message passing is performed by a series of message passing layers, similar to the convolutional layers in a CNN. 

Once the message passing process is finished, the nodes in the graph will have new states, which are a function of their neighboring nodes' states. This new representation can be used for various classes of tasks. There are node-based and edge-based tasks where a node or edge are used to perform a task such as prediction or regression. Since nodes and edges are of a fixed size, traditional machine learning techniques, like an SVM or a ANN, can operate on their representation without any augmentation. There are also graph-based tasks, where the entire graph is used to accomplish a task. As mentioned previously, graphs have no set size or ordering, which is problematic. 

To solve this issue, the idea of ``pooling'' is borrowed from CNNs~\cite{albawi2017understanding}. In GNNs, pooling layers collect the state vectors of all nodes in the graph and collate them into a fixed size vector, which can be used by traditional machine learning techniques. Like message passing layers, there are a variety of pooling layers, such as max pooling, min pooling, and attention pooling, which uses an attention mechanism to learn how to collate state vectors~\cite{velivckovic2017graph}. In recent works, it has been shown that it can be beneficial to use multiple pooling layers in tandem~\cite{corso2020principal,tailor2021we}.

\subsection{Model Interpretability}
 Model interpretability---the process of discerning why a machine learning model makes a certain decision for a given input---is a sought after property amongst machine learning researchers~\cite{petsiuk2018rise,fukui2019attention,dovsilovic2018explainable}. 
Interpretability can help ensure the model is learning to make predictions based on features a domain expert would recognize as important to a given problem.
For example, the tool RISE~\cite{petsiuk2018rise} produces heat maps of images based on how important a pixel is to the model's prediction. 

Ying et al. present a black-box approach to GNN interpretability, GNNExplainer, based on the idea of masking~\cite{ying2019gnnexplainer}. 
Masking is the approach of removing certain data points to see how it affects the model's prediction.
GNNExplainer operates by masking edges in a given graph and giving the altered graph to the model. 
They can then determine the edges that most influence the model's prediction.
\section{Approach}\label{approach}
\subsection*{Problem Statement} 
As SV-Comp has shown, there is no general, optimal program verifier. Selecting the optimal solver requires an in-depth knowledge of the field of verification algorithms and the intricacies of each tool that can be used to prove a specification of a given format. Algorithm selectors give non-expert users of verifiers a way to decide the way in which to run a portfolio of verifiers. We define the problem of algorithm selection for verification as follows:

\begin{definition}\label{def:selection}
    Given a software system, P, a specification, $\phi$, and a suite of verifiers, V, which can accept P and $\phi$, rank V based both on verifiers ability to determine the truth of P $\models \phi$ in terms of correctness and speed, with the former prioritized over the latter.
\end{definition}

Definition~\ref{def:selection} assumes P and $\phi$ are written in a language all verifiers in V can accept. P is commonly written in programming languages such as C~\cite{gadelha2018esbmc,wendler2013cpachecker,dietsch2018ultimate}, Java~\cite{JAVARANGER-SVCOMP20,JDART-SVCOMP20,GDART-SVCOMP22}, and LLVM bytecode~\cite{zhao2012formalizing,legay2020automatic,cadar2008klee}. $\phi$ is written in a specification language, which are typically assertion based~\cite{clarke2006historical} or derived from a temporal logic~\cite{beyer2020advances,jasper2019rers}. Competitions, like SV-Comp, determine the languages of P and $\phi$ that competing verifiers must accept. This results in a set of verifiers which conform to the same programming and specification languages.

\begin{figure*}
    \centering
    \includegraphics[width=\linewidth]{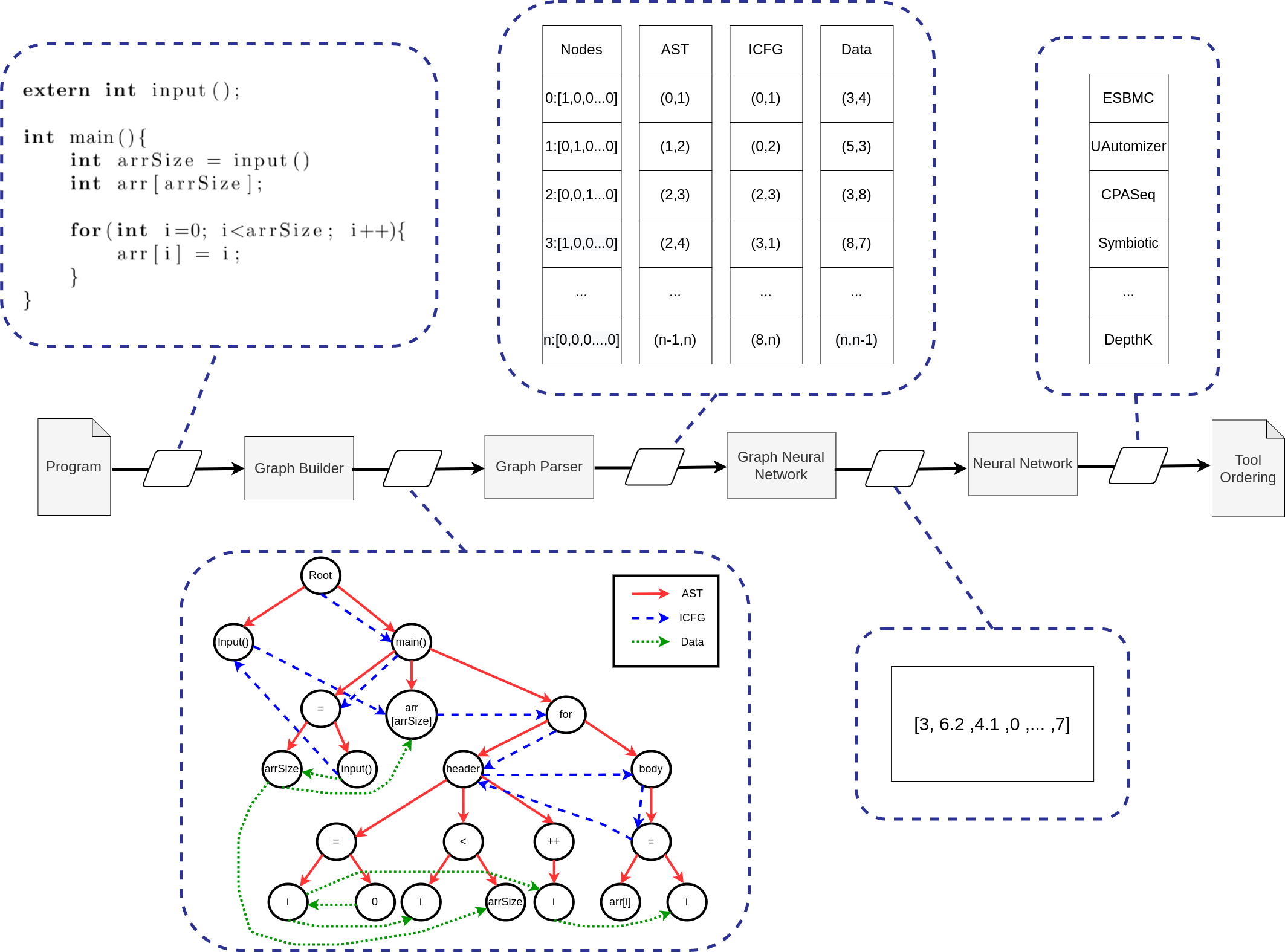}
    \caption{\textsc{Graves} operates using the pipeline shown above. First, it converts programs in graphs by using the program's AST with control flow, data flow, and function call and return edges added between the AST's nodes. It parses this graph into several vectors. These tensors are input to its graph neural network which produces a graph feature vector its simple neural network can make a prediction from.}
    \label{fig:pipeline}
\end{figure*}

\subsection*{Approach Overview}
\textsc{Graves} follows the pipeline shown in Figure \ref{fig:pipeline}. 
Given a C program, it generates a graph representation of the program using the AST as the base with control flow, data flow, and function call and return edges added. 
A parser converts the graph representation into a set of nodes, where node's are represented using a one-hot encoding, and several edge sets. 
A GNN uses these sets to calculate a graph feature vector. 
Finally, a neural network predicts a score for each tool based on its ability to correctly, and quickly verify the given program. 
These scores are then used to rank verifiers from most likely to correctly verify a program to least likely. In the following sections, each component is described in further detail.
\subsection{Graph Generation} \label{graph_gen}
\begin{definition}
We define a graph $G = (\mathcal{V},\mathcal{E},\mathcal{H})$ as follows:
\begin{itemize}
    \item $\mathcal{V} = \{v_0, v_1, ..., v_n\}$ where $v_i$ is a node
    \item $\mathcal{H} = \{h_0, h_1, ..., h_n\}$ where $h_i$ is the one hot encoding of the value of $v_i$
    \item $\mathcal{E} = \{E_0,E_1,...,E_m\}$ where $E_j$ corresponds to the $j^{th}$ edge type in the graph
    \item $E_j = \{e_{j,0}, e_{j,1}, ..., e_{j,p}\}$ where $e_{j,k} \in \mathcal{V} \times \mathcal{V}$ is a directed edge in  $E_j$
\end{itemize}
\end{definition}

For \textsc{Graves}' program graphs, $\mathcal{V}$ is the set of nodes in the abstract syntax tree (AST). 
The AST nodes contain important information about the semantics of a program, such as variables, functions, and operations, but they leave out purely structural tokens, like parentheses or semicolons. 
By construction, the AST retains the information in these structural tokens. 
The node representation, $h_i$, is the one-hot encoding of the AST token associated with $v_i$. 
Let $T$ be the set of possible AST tokens. Each token $t_i\in T$ is represented by an index $i \in [1,|T|]$. 
The one hot encoding of $t_i$, $h(t_i)$, is a vector of length $|T|$ of 0s and a single 1 where $h(t_i) = h(t_j) \Leftrightarrow t_i = t_j$. One hot encodings are often used in graph-based representations for machine learning as they are simple to calculate and have been shown to perform well~\cite{vignac2020building,morris2019weisfeiler,yao2019graph}. 
\changed{In practice, we collect all unique tokens in the training set, $T'$, and map them to a unique integer from 0 to |$T'$|. When a graph is presented to the graph parser, a vector of size $|T'|+1$ is created and the token is passed through the map to receive the location of the 1 in the one hot encoding. Vectors are of size $|T’|+1$ to account for the possibility that during prediction a graph may include a token that was never seen during training.
All such tokens are mapped to the last element of the one-hot vector, which represents the “unknown” token.
}

$\mathcal{E}$ contains three edge sets which capture three types of information: control flow, data flow, and AST edges. Control flow information is vital to most, if not all, verification properties. Control flow will show the path to an error state or the presence of looping behavior which could prevent termination. Thus, it is reasonable to believe that control flow edges are necessary to differentiate between verifiers. For example, techniques which use symbolic execution may perform poorly on loops whose conditions are dependent on program input variables, since they may effectively unroll such loops an arbitrary number of times. Abstract interpretation based techniques can efficiently compute loop fixed points by overapproximating loop condition values to verify properties; this may cost the tool some accuracy. \textsc{Graves} generates control flow edges and function call and return edges using the programs statement based interprocedural control flow graph (ICFG). This graph encodes control flow between each statement in a program and the call and return edges between function calls.

Data flow edges express the way in which variables are used and defined throughout the program. Because \textsc{Graves} abstracts away the names of variables, these edges make explicit the relationship between all definitions of a variable and the uses a definition can reach. Depending on the tool, the formula which describes the value of a variable may affect how its performance. An abstract interpretation based technique may struggle to verify a property which references a variable defined by a nonlinear formula, while a symbolic execution tool should be able to solve the formula to see if it is satisfiable.

Finally, AST edges preserve the structure of the individual statements in a program by connecting operands to operators. 
They can capture complex non-linear expressions which can be problematic for abstract interpreters depending on their abstract domain. 
They are also convenient as every node, besides the root node, is guaranteed to have one AST edge going to it, and in most cases one or more leaving it. 
This allows information not captured in control or data flow edges to propagate through the graph more easily during the message passing phase.

\begin{figure}
    \centering
    \includegraphics[width=\textwidth]{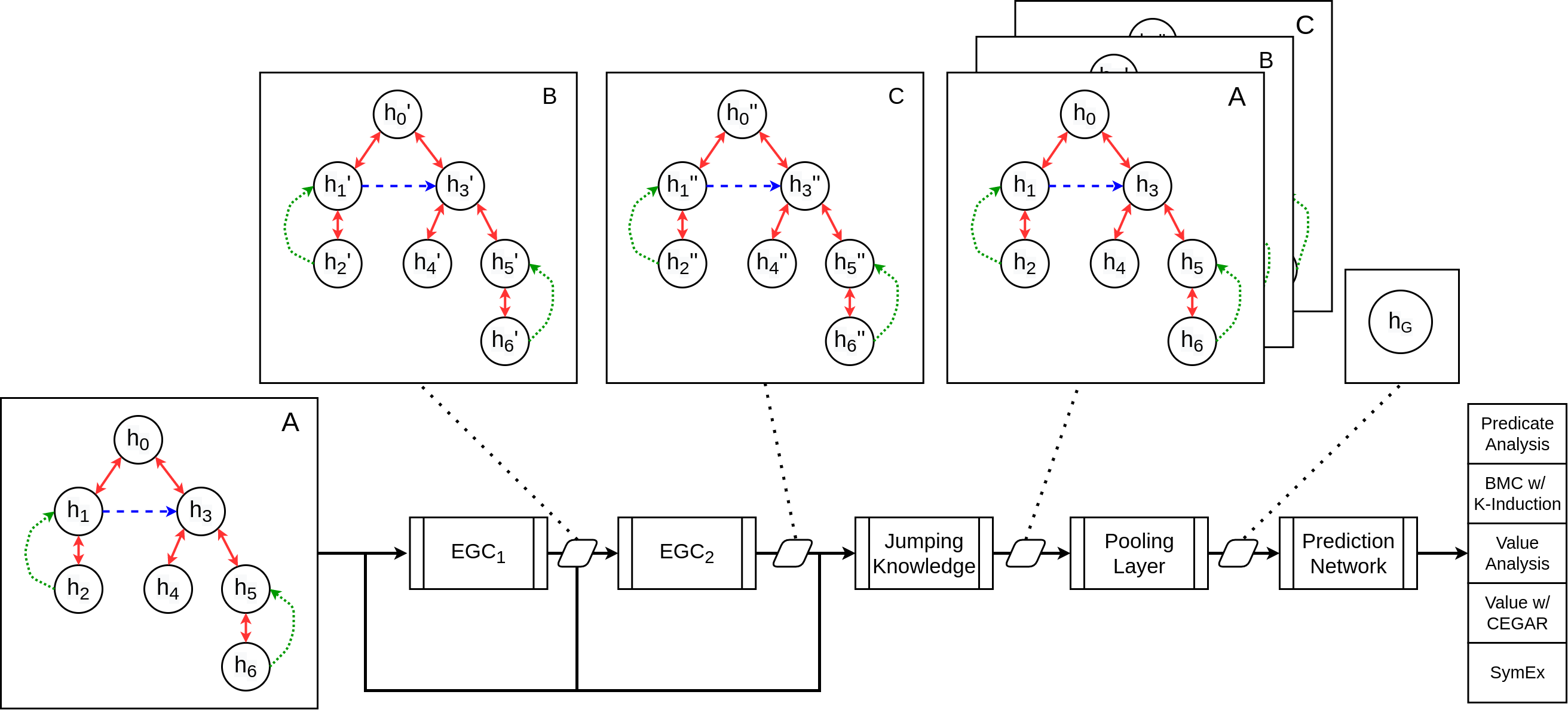}
    \caption{\textsc{Graves} uses a GNN which consists of a variable number of efficient graph convolution layers. A jumping knowledge layer collates the intermediate values of the graph between each EGC layer. The pooling layer merges the nodes in the graph into a single graph feature vector. A fully connected neural network scores each verifier using this vector.}
    \label{fig:gnn_architecture}
\end{figure}
\subsection{Graph Neural Network}\label{GNN}
\textsc{Graves} employs a GNN that consists of three sections: a series of efficient graph convolution layers, a jumping knowledge layer, and a pooling layer.

\subsubsection{Efficient Graph Convolution Layers}\label{EGC}
Many GNN message passing layers are anisotropic, meaning message passing is informed both by the sending and receiving nodes' state. As a result, message passing scales with the number of edges in the graph. As their name suggests, Efficient graph convolutions (EGCs)~\cite{tailor2021we} were introduced to be a more efficient alternative to anisotropic message passing layers. As an isotropic message passing layer, messages are formed only using the state of the node sending the message. Thus, messages can be calculated for each node once and used along every edge said node is passing a message along, allowing message passing to scale with the number of nodes in a graph.

For a node $n_i$ in a graph, the state of $n_i$ after one round of EGC message passing can be formulated as follows. Let $B,H\in \mathbb{N}$, $\mathcal{A}$ be a set of commutative collation functions, $||$ be the concatenation operator, and $\mathcal{N}(i)$ return indices of nodes sending messages to node $n_i$:

\[
    h_{n_i} = \bigg|\bigg|_{h=1}^H \sum_{\oplus \in \mathcal{A}} \sum_{b=1}
    ^B w_{i,h,\oplus,b}\bigoplus_{j\in\mathcal{N}(i)\cup \{i\}} W_b h_{n_j}
\]

For each node $j$ in $n_i$'s neighborhood, $j$'s state vector, $h_j$, is multiplied by a weight matrix $W_b$. These values are aggregated into a single value, $x$, using an aggregation function $\oplus \in \mathcal{A}$. $x$ is multiplied by a learned coefficient $w$. An edge from $n_i$ to itself is created so its state can be incorporated in its state vector's value. This process is repeated $B \times |\mathcal{A}|$ times and summed. It is repeated a further $H$ times, but instead of summation, the resultant values are concatenated. 

H and B are configurable parameters. H is similar to the number of heads in a multi-head attention mechanism~\cite{vaswani2017attention}. B is the number of weight matrices the EGC will learn. EGCs allow node states to be collated using multiple aggregation functions, e.g., mean, max, sum, etc, as previous techniques have found it beneficial to use multiple aggregators to improve accuracy~\cite{corso2020principal}. 

Not only were EGCs found to be more efficient than the state-of-the-art message passing techniques in memory consumption, training time, and inference time, they also achieved significantly higher accuracy on a majority of the benchmarks they evaluated them on. \textsc{Graves} allows for a variable number of EGC layers, as the optimal amount of propagation can be related to the task at hand.

When evaluating \textsc{Graves}, we trained networks using graphs with different sets of edge types. We found that including ICFG and Data edges each had a significant effect on \textsc{Graves} performance. AST edges have a negligible effect on prediction accuracy, but do benefit model interpretability.
The ICFG and Data flow edges that \text{Graves}' graph builder generates are directional edges. Since EGCs are isotropic, messages are formed solely from the node sending information. This mimics control and data flow during message passing.

\subsubsection{Jumping Knowledge Layer} \label{jump}
In image classification networks, it has been shown that early layers in the network can identify coarse features, like the shape of a wheel, and then later layers can identify more fine features, like the spokes in wheel~\cite{tong2017image}.
It is possible that the network can learn from each layer of the GNN. 
\changed{The earlier layers may provide information on local behaviors, such as individual program statements, as they are only a few edge steps away.
Later layers may make calculations on behaviors which take many more steps to find, like the composition of functions.}

Jumping knowledge layers~\cite{xu2018representation} combine the output of several layers, denoted as A, B, and C in Figure~\ref{fig:gnn_architecture}, to produce an aggregate representation, typically using concatenation, max-pooling, or a recurrent layer. In \textsc{Graves},
graph representations produced by the intermediate EGC layers are fed to the jumping knowledge layer which concatenates the intermediate and final representations of each node into a single vector.
This allows \textsc{Graves}' network to learn on the intermediate node representations along with the final representation produced by the EGC layers.

\subsubsection{Pooling Layer} \label{pool}
Graphs must ultimately be collated into a fixed sized representation to perform the task of graph prediction. 
Recent work has shown that using several pooling functions and combining their outputs into a single vector produces better performance than a single pool and adds minimal overhead~\cite{corso2020principal}.
\cite{tailor2021we} suggests that pools chosen be diverse in what they calculate.

There are three types of pooling operators: collation pools, representative pools, and learned pools. Collation pools combine the state vector of each node in the graph, e.g., sum, mean, product, etc. For each dimension $i$ in the state vector, representative pools select a single node's value at dimension $i$ to represent the graph, e.g., min, max, median, etc. Learned pools, such as equilibrium~\cite{bartunov2022equilibrium}, attention~\cite{velivckovic2017graph}, or graph multiset pools~\cite{baek2021accurate}, evolve during the training process and learn a function which can combine the state vectors in a way that fits the problem at hand.

We evaluate \textsc{Graves} in Section~\ref{model_evaluation} with mean, max, and attention pooling operations, representing the collation, representative, and learned pools respectively. Mean pooling helps the GNN learn representations which capture the distributions of features across the entire graph~\cite{xu2018powerful}. A mean pool may cause the GNN to capture the distribution of nonlinear path conditions in a program as tools based on abstract interpretation can struggle in these situations, while symbolic execution based tools will not. Max pooling has been found to help the model learn representations which capture distinct elements~\cite{xu2018powerful}. A max pool may cause the GNN to identify unbounded loops as bounded model checkers will often struggle to verify programs with them, but bounded model checkers using k-induction are better equipped to handle them. Attention pools are a generalization of a sum pool. In a sum pool, the attention value is always 1. Sum pools have been shown to enable to network to learn representations which capture the structure of the graph~\cite{xu2018powerful}. The attention mechanism will learn to amplify the values of important graph structures and minimize the value of unimportant ones when performing summation.

The output of each pool can be combined using some of the very same functions as the pools themselves, i.e., mean, max, attention, etc. \textsc{Graves} concatenates the outputs of the pools as this is the default and performs well.

\changed{
\subsection{Property Representation}
In verification, $\phi$ can be categorized into different property classes. Prior work~\cite{beyer2021software,beyer2022progress} has defined the following classes:
\begin{itemize}
    \item Reach Safety -- an ``error state'' is not reached
    \item Memory Safety -- freeing memory and pointer dereferencing is done safely
    \item No Overflows -- integer overflow cannot occur
    \item Termination -- the program with always terminate
\end{itemize}
Of these classes, only Reach Safety properties are explicitly stated in the program, via user defined assertions or error statements. The remaining classes of properties are implicit. Overflow checks must occur anywhere in the program integer values are manipulated. To prove termination, the verifier must determine if there is looping or recursive behavior in the program and that it will eventually cease.

Previous algorithm selection techniques suggested training a selector for each property class, as they saw improved accuracy. This is likely because the selector didn't incorporate the property class into their prediction. For a program P and two verifiers $V_A$ and $V_B$, it is possible that $V_A$ could prove P terminates but cannot determine if an error state is unreachable, while $V_B$ can do the opposite~\cite{beyer2022progress}. If a selector makes predictions based only on the program, as prior selectors do, it will select the same verifier for the program even if the property changes.

Unlike prior techniques, \textsc{Graves} incorporates the property class into its prediction, allowing it to make different predictions for the same program based on the property. \textsc{Graves} maps each property class to a single integer---i.e., reach safety to 0, memory safety to 1, etc. This value is then appended to the output of the GNN pooling layer, $h_G$. This approach, while simple, is effective. During training, \textsc{Graves}' GNN learns how to augment its representation of the program and its final prediction based on the property class value.

\subsection{Prediction Network}\label{ffnn}
$h_G$ is a vector with a fixed size, which means it can be passed through \textsc{Graves}' prediction network.
\textsc{Graves} uses a simple three layer fully connected neural network. 
This network produces a score for each verifier in the portfolio based on how likely they are to verify $P$ and how fast. 
Using these scores, we can rank the verifiers from most effective to least effective, in terms of correctness followed by speed.
}

\subsection{Implementation} \label{implementation}
We created an implementation of the \textsc{Graves} approach for C programs and C program Verifiers which can be found in our GitHub repository\footnote{https://github.com/will-leeson/graves}. It contains 1306 SLOC in C++ and Python. Our implementation is highly configurable, allowing users to vary the number of EGC layers, EGC's variable components (H, B, $\mathcal{A}$), the number and types of global pooling operations (attention, max, mean, variance, etc.), and the training hyperparameters (number of epochs, learning rate, learning rate scheduling, etc.). We provide scripts to replicate our study.

To create program graphs, \textsc{Graves} uses the AST generated by the C compiler Clang~\cite{clang}. 
Using a visitor pattern~\cite{gamma1995design}, it walks the AST to collect its nodes and edges. 
\textsc{Graves} also collect the information that is necessary to generate control, call, return, and data flow edges. 
Using a work-list reaching definition algorithm~\cite{aho2007compilers}, \textsc{Graves} generates data flow edges.

We implement \textsc{Graves}' GNN and neural networks using the machine learning library PyTorch~\cite{NEURIPS2019_9015} and an extension of the library, PyTorch Geometric~\cite{Fey/Lenssen/2019}. 
PyTorch Geometric is a machine learning framework made to perform deep learning on graph and irregularly shaped data. 
It has implementations of many state-of-the-art GNNs techniques as well as a method to create new layers. 
\section{Related Work}\label{related}
In this section, we describe uses of graph neural networks to perform software engineering tasks, existing approaches to algorithm selection for verification and compare said approaches to \textsc{Graves}.

\subsection{Graph Neural Networks for Software Engineering} \label{GNNSE}
Graph neural networks have been used to represent programs for various purposes in the software engineering community.
Typically, techniques begin with the program's abstract syntax tree. 
From there, they add edges representing information they find may be useful in solving their task. 
Tasks that focus on individual statements may add edges between the tokens in statements in the order they appear, allowing the network to create statement based feature vectors. 
Tasks focused on program optimization may require edges between function calls or variable uses to more accurately summarize the entire program.

GNNs are capable of performing tasks to assist programmers in developing software. 
Allamanis et al.~\cite{allamanis2017learning} use graph neural networks to perform tasks akin to a linter~\cite{johnson1977lint}. 
They find when a variable has been used incorrectly in place of another and they predict the names of such variables. 
These tasks can be integrated into IDEs to prevent misuses of variables and to help developers give descriptive names to variables, making source code more legible. Both LeClair et al.~\cite{leclair2020improved} and Lu et al.~\cite{lu2019program} use graph neural networks to classify programs. 
This can be helpful for code mining and automatic code completion.

GNNs can also be used for tasks dealing with resource management. 
Cummins et al.~\cite{cummins2020programl} use graph representations of programs and GNNs to perform compiler analysis tasks, such as determining the sequence in which optimizations should be applied. This has the potential to improve both compilation time and the execution time of programs.

To perform message passing, Allamanis et al., Cummins et al., and Lu et al. use Gated Graph Neural Networks (GGNNs) and LeClair et al. use Convolutional Graph Neural Networks. Recent work has introduced new message passing layers which, in general, outperform them~\cite{velivckovic2017graph,tailor2021we}. They each also use different graph structures. LeClair et al. only use the program's AST. While the AST does maintain program structure, it lacks information such as loop return edges or def/use pairs. On the other extreme, Allamanis et al. use 10 different edge types, which may allow unnecessary information to propagate through the graph. \textsc{Graves}' graph builder is configurable and can contain the program's AST, control flow edges, data flow edges, and function call and return edges. Many program verification tools use program graphs based on this same information.

\subsection{Algorithm Selection for Verification} \label{selectionRelated}

While machine learning is the norm, it is not required to perform algorithm selection for verification. 
The tool VeriAbs~\cite{darke2018veriabs} implements an algorithm selector which uses a rule based system. 
Using lightweight analyses, they compute information about loops and the inputs to programs to decide which verification techniques to apply to the program. 
This technique proved to be very successful and VeriAbs has won the ReachSafety category in SV-Comp for the last four years.

To create a machine learning based algorithm selector for program verification, programs must be represented in a way amenable to machine learning techniques. 
Raw text is a poor representation of programs. 
Natural language processing has been used to ``understand'' a program's raw source code to perform some tasks~\cite{ernst2017natural,allamanis2018survey}, but programs are more than just a sequence of statements. 
The order in which statements are executed is based on the branching behavior of a program. 
Another way to represent a program is by extracting statistics, such as counts of loops, pointers, or occurrences of recursion, to form a ``feature vector''. 
Models have been trained using these feature vectors to select verifiers with some success~\cite{demyanova2017empirical,tulsian2014mux}. 
This representation will capture the presence of certain behaviors in a program, but not how they interact with each other. 
If program A has a loop with an assertion in it and program B has a loop with an assertion before the loop, counting these features would show no difference between A and B. 
This is problematic as some algorithms falter with certain looping behaviors, while other excel.

Similar to \textsc{Graves}, WLJ~\cite{richter2020algorithm} operates on graph representations of programs. They calculate new graph representations by concatenating the label of adjacent nodes for a set number of iterations.
They then use an SVM which compares subgraphs in these representations using the Weisfeiler–Lehman test for graph isomorphism~\cite{shervashidze2011weisfeiler}. WLJ and \textsc{Graves} make use of similar graph representations of programs. The main difference between the two techniques is how they operate on their generated graphs. The process of concatenating node representations does not take into account the importance of various tokens to the given problem. \textsc{Graves}'s EGC layers learn how to combine node representations, which allows it to tailor representation propagation to the given problem. Further, WLJ's SVM operates on subgraphs of the final graph representation. This can be problematic when aspects of the graph which affect a tools ability to verify a problem are further than the subgraph size limit. \textsc{Graves}' global pooling mechanisms allows it to form a final representation of the graph which is informed by every node in the graph. 

The authors of WLJ implemented their technique in a tool called PeSCo~\cite{richter2019pesco} which has competed in SV-Comp since 2019. PeSCo is a fork of the CPAChecker framework~\cite{beyer2011cpachecker} which they have integrated the WLJ selector into. PeSCo selects from 6 configurations of CPAChecker which can solve problems in the reach safety category. For all other specifications, they use a single, default CPAChecker configuration. PeSCo has performed well in the competition, placing between $2^{\texttt{nd}}$ and $4^{\texttt{th}}$ in the overall category every year it has participated.

In preliminary work, we created a prototype of \textsc{Graves}, \textsc{Graves-CPA}, which also selects from configurations of CPAChecker. We entered \textsc{Graves-CPA} into SV-Comp 2022 and it came in $5^{\texttt{th}}$ place overall, just behind PeSCo. PeSCo performed better than \textsc{Graves-CPA} in three categories---Reach Safety, Falsification Overall, and Overall---and \textsc{Graves-CPA} beat PeSCo in the Software Systems category. \textsc{Graves-CPA} uses a simpler GNN architecture than the one we proposed in Section~\ref{GNN} and a CPAChecker configuration file that only allowed for a subset verification approaches used by PeSCo to be predicted. In Section~\ref{RQ1}, we perform a head-to-head comparison between PeSCo and an updated version of \textsc{Graves-CPA} which uses the updated GNN architecture and the same configurations as PeSCo. This will isolate the comparison to the selectors PeSCo and \textsc{Graves-CPA} use, as it is the only difference between the tools.

In Richter et al.~\cite{richter2020attend}, they teach a network to encode a version of the abstract syntax tree of programs. \textsc{Graves} also makes use of the abstract syntax tree, however, the graph is enriched with control and data flow information. This allows information to flow across control and data dependencies. They encode graphs using an attention network, which allows it to weight encodings when propagating information, unlike PeSCo's static propagation. Their graph structure limits how information propagates, only allowing it to flow up the tree to the root node. \textsc{Graves}' GNN allows information to propagate through the graph in either direction. This allows for each operand's encodings to be informed by the operator associated with them, different functions to learn about each other, etc.
\section{Research Questions and Experimental Design}\label{design}
To evaluate \textsc{Graves}, we answer the following questions:
\begin{description}
    \item[\textbf{RQ1}]{How does \textsc{Graves} compare to other algorithm selectors for verification?}
    \item[\textbf{RQ2}]{How does using \textsc{Graves}' improve verification in practice?}
    \item[\textbf{RQ3}]{How do the components of the \textsc{Graves}' GNN architecture affect its predictions?}
    \item[\textbf{RQ4}]{Does \textsc{Graves} identify program patterns associated with verification algorithm success?}
\end{description}


\subsection{Baseline techniques}
To address \textbf{RQ1}, we select several baseline techniques to compare against \textsc{Graves}. The selector introduced in~\cite{richter2020algorithm} produces program graphs similar to \textsc{Graves}. However, they do not use graph neural networks to select verifiers. As described in Section~\ref{selectionRelated}, they introduce an SVM with a specialized kernel to look for similarities in graphs. This SVM uses the Weisfeiler-Lehman test to calculate Jaccard similarity, so we refer to this technique as WLJ.

The selector introduced in~\cite{richter2020attend} introduces a variant of ASTs called contextualized syntax trees (CSTs). In CSTs, nodes of the AST are grouped in hierarchies, e.g. function, statement, token, etc. We refer to this technique as CST.

We evaluate two additional selectors: an ideal static selector (ISS) and a random selector. The purpose of ISS is to evaluate the benefits of dynamic selection. If the dynamic selectors perform similarly to ISS, then the benefit of the dynamic selector may not outweigh the cost of its overhead. The random selector allows us to evaluate our metrics. If a random selector performs well for a given metric, then the metric is most likely not rigorous.

\subsection{Verifier suites}
In order to evaluate \textsc{Graves} and the baseline techniques, we must instantiate a suite of verifiers, V, for each technique to select from.

For the evaluation of \textsc{Graves}, WLJ, and CST in \textbf{RQ1}, \textbf{RQ2}, and \textbf{RQ3}, the suite of verification tools are the SV-Comp 2018 submissions of the following tools: 2LS~\cite{schrammel20162ls}, CBMC~\cite{kroening2014cbmc}, CPA-Seq~\cite{wendler2013cpachecker}, DepthK~\cite{rocha2017depthk}, ESBMC-KInd~\cite{gadelha2018esbmc}, ESBMC-Incr~\cite{gadelha2018esbmc}, Symbiotic~\cite{chalupa2017symbiotic}, Ulitimate Automizer~\cite{heizmann2013software}, Ultimate Kojak~\cite{nutz2015ultimate}, and Ultimate Taipan~\cite{dietsch2018ultimate}. 
The labels for these verifiers come from the results reported at SV-Comp 2018~\cite{sv-results}. 

For the evaluation of \textsc{Graves-CPA} and PeSCo in \textbf{RQ1}, the tools select from six verification algorithms implemented in the CPAChecker framework: bounded model checking (BMC)~\cite{biere1999symbolic}, bounded model checking with K-Induction (BMC+K)~\cite{de2003bounded}, predicate analysis~\cite{graf1997construction}, value analysis~\cite{cousot1977abstract}, value analysis with counter-example guided abstraction refinement (CEGAR)~\cite{clarke2000counterexample}, and symbolic execution (SymEx)~\cite{king1976symbolic}. These are the same 6 configurations PeSCo selected from in SV-COMP 2022. CPAChecker allows for configurations to be used conditionally. If recursion is detected, both PeSCo and \textsc{Graves-CPA} default to Block Abstraction Memoization~\cite{wonisch2012predicate}.

For the evaluation of \textbf{RQ4}, \textsc{Graves} selects from 4 verification algorithms implemented in the CPAChecker framework: BMC, BMC+K, CEGAR, and SymEx. 
We select BMC, CEGAR, and SymEx as they are distinct techniques with separate benefits and shortcomings. 
We include BMC+K to observe if the network can identify the advantages and disadvantages K-Induction adds to BMC. 
Since \textsc{Graves} is selecting from tools within one framework, it avoids the issue of identifying implementation specific details as a reasons for selection, like supporting various data types. It also ensures \textsc{Graves} is selecting from individual techniques as most tools use an amalgam of techniques.

\subsection{Datasets} \label{Dataset}
To train and evaluate the \textsc{Graves} and the other machine learning based techniques, we must provide them a large set of verification problems and divide them into train, evaluation, and test sets.

SV-Comp evaluates tools built for both the C and Java programming languages.
As the Java competition is relatively new, the set of benchmarks and tools is not as diverse as those in the C competition.
SV-Comp has a large set of verification problems with several possible specifications written in C. 
In their evaluations, WLJ and CST used the 2018 SV-Comp benchmarks and selected 10 verifiers which competed in all four major categories: reach safety, termination, memory safety, and overflow.~\cite{SV-Benchmarks2018}. Definitions of these specifications can be found on the SV-Comp website\footnote{https://sv-comp.sosy-lab.org/2018/rules.php}. 

The 2018 SV-Comp benchmarks consists of 9523 verification problems written in C.
We remove roughly 13\% of the examples from the data set as they produce graphs too large for training given the constraints of the GPUs we have available for experiments (VRAM is limited to 16GB).
For reference, the average graph from the remaining 8330 examples is roughly 8.75 MB. This is purely a training limitation. Training requires gradient computation, which scales poorly with graph size. During evaluation when gradient computation is turned off, \textsc{Graves} can operate on large graphs and make predictions.

\textsc{Graves-CPA} and PeSCo were trained on the 2021 SV-Comp Reach Safety benchmarks and evaluated on the 2022 SV-Comp Reach Safety benchmarks in competition. In \textbf{RQ1}, we evaluate \textsc{Graves-CPA} with our new GNN architecture. For fairness, we train this new version of Graves-CPA on the same benchmarks as our 2022 competition contribution. To evaluate \textsc{Graves-CPA} and PeSCo's ability to generalize, we evaluate them on the 2023 SV-Comp benchmark set, the newest set at the time of writing this. The 2021 SV-Comp Reach Safety benchmark set consists of 8,452 problems and the 2023 SV-Comp Reach Safety set adds an additional 1,459 problems.

For each dataset, the label is the verifier's SV-Comp score for the given benchmark \textit{b} minus a penalty for the time it took to verify \textit{b}. 
We randomly divide our data into training, validation, and test sets. SV-Comp requires that machine learning based techniques train on no more than 20\% of the benchmark in an effort to prevent the technique from memorizing the benchmarks. To train \textsc{Graves-CPA} for our SV-Comp 2022 and the evaluation of \textbf{RQ1}, we randomly selected 20\% of the 2021 SV-Comp Reach Safety benchmarks and divide them into training and evaluation sets using a 80-20 split. The test set is the entire SV-Comp 2023 Reach Safety benchmarks. For all other evaluations, we randomly divided the 2018 SV-Comp benchmarks into train, evaluation, and test sets using an 80-10-10 split. We ensure that the split reflects the populations of specifications, i.e., 80\% of reach safety problems are in the training set, 10\% are in the validation set, and 10\% are in the test set. 

\subsection{Network Training}
To train CST models, we looked to the repository listed in their paper to replicate their results. After interacting with the authors, we could not reproduce the results of WLJ. As a result, we omit WLJ from our evaluation of successful verifier accuracy and Top-K accuracy as they did not evaluate their technique using these metrics. For Spearman correlation, we quote the results from their paper~\cite{richter2020algorithm}. In \cite{richter2020algorithm}, the authors list several configurations of WLJ, parameterized by two values, i and j. The i value corresponds to the number of node relabeling iterations they complete, similar to message passing. The j value corresponds to the depth of the AST their selector looks at when comparing graphs. We report the optimal selector from their study, where $\text{i}=0$ and $\text{j}=5$.

There are several configurable parameters of \textsc{Graves}: the edge sets included in the program graph, the number of EGC layers, the H, B, and $\mathcal{A}$ parameters of the EGC layers, and the set of pooling operations in the pooling layer. For each EGC layer, H and B are set to their default values, 8 and 4 respectively. For the remaining parameters, we performed a grid search~\cite{liashchynskyi2019grid} of these parameters. We found $\mathcal{E}$ = \{AST, ICFG, Data\}, 2 EGC layers with $\mathcal{A} =$ \{max, mean, std\}, and a pooling layer consisting of max, mean, and attention pooling operations to be the best configuration. All results use the aforementioned configuration, unless stated otherwise. Note that it is possible to achieve better performance for specific values of $\phi$ with different configurations.

For \textsc{Graves}' model, we performed hyper-parameter tuning, varying epochs among $\{ 25, 50, 100 \}$ and learning rate among $\{$1e-3, 1e-4, 1e-5$\}$. 
We found a learning rate of 1e-3 and 25 epochs to be optimal.
We train in batches of size 5, as this allowed networks to reliably finishing training without running out of GPU VRAM due to our constraints.
We used a learning rate scheduler to train \textsc{Graves}' network.
After three consecutive epochs where the network's validation loss did not improve, the learning rate was decreased one order of magnitude. 
Networks were trained for the chosen number of epochs, or until the learning rate fell to 1e-8. 
\textsc{Graves} uses a pair-wise margin ranking loss to train networks as this penalizes poor ranking of any verifier, regardless of its true position in the label.

\subsection{Training and Evaluation Environments}
Training took place on various machines with different specifications. 
Aside from the limitation on graph sizes imposed by GPU VRAM, training resources are not pertinent to any of our research questions, so this poses no threat to validity.

In our evaluation of PeSCo and \textsc{Graves-CPA} on the SV-Comp 2023 benchmarks for \textbf{RQ1}, we evaluate the tools on identical Centos machines with one 2.50GHz Intel(R) Xeon(R) Gold 6248 CPU. We attempted to simulate the SV-Comp 2023 environment as closely as possible, including validating the witnesses produced by each tool. The tools where executed in Singularity Containers~\cite{kurtzer2017singularity} running Ubuntu 22.04. Verification runs had access to 8 CPU cores, 15GB of RAM, and given a 15 minute maximum runtime. Validation runs had access to 2 CPU cores and 7GB. For true problems, validators where given a maximum runtime of 15 minutes. For false problems, validators where given a maximum runtime of two minutes. We ran 4 validators which participate in SV-Comp: CPAChecker~\cite{beyer2016correctness}, CPA-w2t~\cite{beyer2018tests}, Symbiotic-Witch~\cite{ayaziova2022based}, and Ultimate Automizer~\cite{beyer2018tests}. 

For \textbf{RQ2}, we evaluate the overhead of selectors on identical CentOS servers. Each server has one 2.40GHz Intel(R) Xeon(R) E5-2680 CPU, 32 GB of RAM, and 1 Nvidia P100 GPU with 16 GB of VRAM. For \textbf{RQ4}, labels were collected by running each CPAChecker configuration on the SV-Comp 2018 dataset on identical CentOS servers with one 2.10GHz Intel(R) Xeon(R) Gold 6130 CPU and 128 GBs of RAM. 

\subsection{Evaluation Metrics} \label{metrics}
We evaluate \textsc{Graves} against the baseline techniques using three metrics: the ability to predict a successful verifier, Spearman rank correlation~\cite{spearman1987proof}, and Top-K error. The first two metrics were used in the evaluation of \cite{richter2020attend} and \cite{richter2020algorithm}, respectively. 

\subsubsection{Successful Verifier Selection Accuracy} \label{success}
The simplest measurement we evaluate \textsc{Graves} on is the ability to select a verifier which will be successful in verifying a given program. 
We remove instances where no verifier could solve the given instance from the test set when evaluating this metric as this would artificially deflate the results. 
We also remove instances where all verifiers could solve the given problem as this would artificially inflate the results since any response is correct.
We are left with 680 examples, or 81.8\% of the test set.
For this metric, ISS is the verifier which produces the most correct responses on the training set.

\subsubsection{Spearman Rank Correlation} \label{spearman}
Spearman rank correlation~\cite{spearman1987proof} determines how similar two lists are ordered, where $\rho(x,y)=1$ implies x and y are ordered them same and $\rho(x,y)=-1$ implies $x$ is the inverse order of $y$. In our case, x is the true score of each verifier and y is the predicted score for each verifier.
It is closely related to Pearson correlation coefficient which measures the linear correlation between two lists~\cite{pearson1896vii}. 
The Spearman rank correlation coefficient of two lists, x and y, is equal to the Pearson correlation coefficient of the ranking of x and y. 
Spearman rank correlation is defined as follows, where X and Y are lists of n values and $x_i$ and $y_i$ are the rankings of the $i^{th}$ values in X and Y, respectively:
\[
    \rho(X, Y) = 1 - \frac{6 \sum_{i=1}^{n}(x_i - y_i)^2}{n(n^2-1)} \in [-1,1]
\]

The ISS ranks verifiers using the Borda counts~\cite{Behnke2004} on the training set. 
The Borda count for each verifier $v$ can be calculated as follows, where n is the number of verification instances, k is the number of verifiers in our suite, and $r_i$ is the true ranking of the verifier on verification instance $i$:
\[
    \texttt{Borda(v)} = \sum_{i=1}^n k - r_i
\]
Borda counts are the optimal static ordering for any ranking in terms of Spearman rank correlation~\cite{hullermeier2010predictive}. 

Spearman rank correlation is interesting in the case where verifiers are run in sequence. A network with high Spearman rank correlation should choose the verifiers most likely to succeed in order of speed of verification. 

\subsubsection{Top-K Error} \label{topk}
Top-K error is a metric often used to evaluate deep learning models on the task of object recognition in images~\cite{krizhevsky2012imagenet,ren2015faster,he2016deep}. 
Given some k-value, if the label value is within the first k choices the network predicts, then error is 0. 
If the label is not in the first k choices, error is 1. 
In our case, the label is the verifier which performs best, in terms of accuracy and then time, on the given program.

The ISS chooses tools in order of how often they were the best selector on the training set, meaning the tool which was the best tool most often is first, the tool which is best the second most often is selected second, etc. 
This guarantees the highest Top-K values for a static ordering of tools on the training set.

A critique of algorithm selection is that a user could run the portfolio of algorithms in parallel instead of selecting one at a time~\cite{kerschke2019automated}. 
This has the potential to cause an excessive use of resources as verification techniques tend to scale poorly in both time and memory usage as program complexity grows. 
While full parallelization is guaranteed to verify a program in the shortest time for the set of verifiers, there will be a waste computation time and energy on all verification tools that do not report a safety guarantee first. 

Algorithm selectors can be used to find a middle ground between parallelism and sequential selection by training a selector to predict the Top-K tools most likely to verify a program, where k is the level of parallelism. 
If the selector can consistently pick the most effective tool in k choices, the developer will receive a safety guarantee in the same amount of time, while using a fraction of the computational resources.
\section{Model Evaluation} \label{model_evaluation}
In the following section, we discuss our model's performance across a variety of metrics. We look at how it compares to previous techniques and how different components of the network affect its ability to make predictions.

\subsection{Model Performance} \label{RQ1}
\begin{figure*}
\begin{tabularx}{\linewidth}{*{2}{>{\centering\arraybackslash}X}}
\resizebox{1\linewidth}{!}{
\begin{tabular}[b]{@{}lrr@{}}
\toprule
\multicolumn{1}{c}{\textbf{Algorithm Selector}} & \multicolumn{1}{c}{\textbf{Success}} & \multicolumn{1}{c}{\textbf{Spearman}} \\ \midrule
\textsc{Graves} & \textbf{0.856 $\pm$ 0.007} & \textbf{0.734 $\pm$ 0.001} \\ \arrayrulecolor{gray}\midrule
WLJ & - & 0.654 $\pm$ 0.014 \\ \midrule
CST & 0.577 $\pm$ 0.000 & 0.257 $\pm$ 0.000 \\ \midrule
ISS & 0.697 $\pm$ 0.000 & 0.243 $\pm$ 0.000 \\ \midrule
\arrayrulecolor{black}
Random & 0.414 $\pm$ 0.009 & 0.002 $\pm$ 0.012 \\ \bottomrule \\~\\
\end{tabular}}
&
\includegraphics[width=\linewidth]{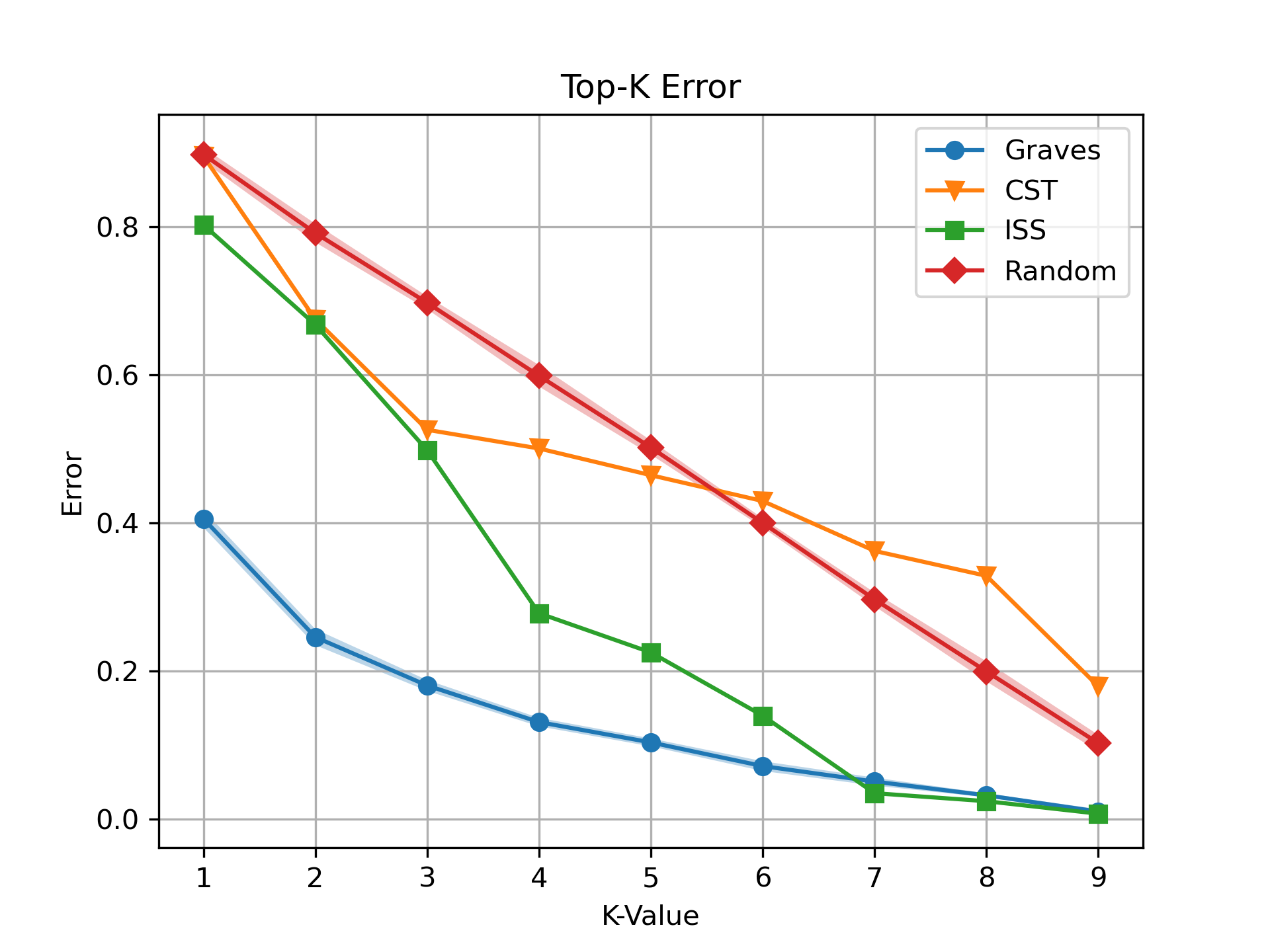}
\end{tabularx}
\captionof{figure}{\textsc{Graves} is the best selector across all three metrics. CST is able to select a successful verifier more than half of the time, but is actually worse than the static selector.
The fact that the Random selector perform as well as it does at the success verifier selection shows how easy of a task it is. 
WLJ was only evaluated on the task of ranking, but it performs well at it, only roughly 12\% worse than \textsc{Graves}. 
For Top-K error, \textsc{Graves} is the best performing technique until K=7 where it performs similarly to the ideal static ordering. 
When K$>$5, if a verifier were randomly selected, the best verifier is more likely to be selected than not.}
\label{fig:overallResults}
\end{figure*}

We address \textbf{RQ1} by comparing \textsc{Graves} against the baseline techniques across the metrics we discuss in Section~\ref{metrics}. 
Figure~\ref{fig:overallResults} lists the results for the Successful Verifier Selection Accuracy, Spearman Rank Correlation, and Top-K Error.
For Successful Verifier Selection Accuracy and Spearman Rank Correlation, we list the average and standard deviation of the results of 10 selectors of each technique. 
For Top-K Error, we present a line graph where each point is the average Top-K Error for 10 selectors of the given technique. 
Above and below each line is a shaded region which represents one standard deviation from the average. 
Note that for many techniques this is barely perceptible as their standard deviation is very low.
The CST networks have minuscule standard deviation ($<1e^9$) due to the fact that they use a static seed in training their networks. 
ISS has no standard deviation as it is deterministic.

\subsubsection*{Successful Verifier Selection Accuracy} 
\textsc{Graves} shows a 23\% improvement, or a 16 percentage points increase, over the next best technique, ISS, and a 48\% improvement, or a 28 percentage point increase, over the best performing selector (CST). 
The CST model does not take specification type into account, which \textsc{Graves} does. 
Both CST and \textsc{Graves} use a simple feedforward neural network to make final predictions. 
It is reasonable to believe that CST could perform better if it encoded $\phi$ and append it to their feature vector, like \textsc{Graves} does.

In general, the problem of selecting a successful verifier is not very interesting. Simply by randomly choosing a verifier, there is roughly a 40\% chance of making a correct choice. Because verification can be such an expensive process, it is important to select not only a verifier that can verify a system, but one that can do it efficiently.

\subsubsection*{Spearman Rank Correlation}
\textsc{Graves} improves on the state-of-the-art (WLJ) by 12\%, or 8 percentage points, in terms of Spearman rank correlation. The ideal static selector performs much worse on this metric as there is actual competition as to which verifier performs best on a given instance. This is a more rigorous metric than successful verifier selection accuracy as it takes into account the quality of the tools performance and not solely if it could accomplish a task. Tools are penalized according to how long it takes to verify a program.

An issue with Spearman rank correlation is that it rewards getting the last item correct just as much as getting the first item correct. Let label = $[0,1,2,3,4,5,6]$, pred$_a = [0,1,2,6,4,5,3]$, and pred$_b = [3,1,2,0,4,5,6]$. $\rho(\text{label}, \text{pred}_a) = \rho(\text{label}, \text{pred}_b)$. Since pred$_a$ predicts the best verifier first, it is a better ordering than pred$_b$. Selectors may be receiving high scores simply because they can order poorly performing verifiers better. In order to get faster and more accurate verification results, we want selectors to predict the best verifiers for a given problem first. It is less important that they get the correct ordering for the verifiers which will perform poorly. 

\subsubsection*{Top-K Error}
\textsc{Graves} is decidedly better than any other selector until K$>$6. After this, the difference between \textsc{Graves} and ISS is negligible. Once again, we omit WLJ as we could not replicate their studies to collect the appropriate data. We argue that K$>$5 is not an interesting metric, as there are 10 verifiers to select from. Randomly selecting more than 5 verifiers is going to include the best verifier more often than not. 

These results can help us infer some interesting information about previous metrics. The Top-1 error shows how often the technique selects the best performing verifier. This implies that when \textsc{Graves} chooses a ``successful'' verifier, roughly 60\% of the time, it is the best performing verifier. CST on the other hand chooses the best performing verifier only approximately 15\% of the time. These results also show that the best performing verifier is one of the first 3 verifiers \textsc{Graves} selects over 80\% of the time. This suggests that the network is not succumbing to the Spearman rank correlation issue mentioned earlier of optimizing the order of the poor performing verifier and ignoring the order of the best performing ones.

We argue Top-K error is a better metric than Spearman correlation or successful verifier accuracy to measure algorithm selectors. 
Moreover, Top-K error reveals how much parallelization is needed on average to reach the best result.
On 87\% of the problems in the test set, the solvers \textsc{Graves} selects to run on a 4-core CPU will be able to achieve the best performance, i.e., since the top-performing verifier will be among the Top-4.

\subsubsection*{Category Specific Training}
So far, we have evaluated the algorithm selectors when $\phi$ is variable.
In~\cite{richter2020algorithm}, the authors find the WLJ performs better when $\phi$ is fixed. As described in Section~\ref{ffnn}, \textsc{Graves} can and does incorporate $\phi$ into its predictions, but WLJ is not easily able to, and CST can, but does not. As a result, \textsc{Graves} often gains very little from training on problems of a single specification as opposed to training on the entire set of problems. When trained only on termination or overflow problems, \textsc{Graves} improves roughly 5\%. When trained exclusively on reach safety problems, \textsc{Graves} improves less than 0.2\%. When trained solely on memory safety problems, \textsc{Graves}' performance actually decreases nearly 4\%. When trained exclusively on reach safety or overflow problems, it is possible for WLJ to outperform \textsc{Graves} in terms of Spearman correlation for problems of these specifications. See the Appendix for an evaluation on each specification in our dataset. 

\begin{table}[]
    \caption{SV-Comp 2023 Simulation. On all four categories they compete in, we find that \textsc{Graves-CPA} outperforms PeSCo. Perhaps most importantly, \textsc{Graves-CPA} performs better on Software System problems. This implies that \textsc{Graves} generalizes better to realistic software developers interact with daily.}
    ~\label{tab:SVCOMP23}%
        \begin{tabular}{@{}lrrrr@{}}
        \toprule
        Tool                & \multicolumn{1}{l}{Reach Safety} & \multicolumn{1}{l}{Software Systems} & \multicolumn{1}{l}{Falsification} & \multicolumn{1}{l}{Overall} \\ \midrule
        PeSCo               & 4763                             & -470                                 & 45                                & 3517                        \\ \hline 
        \textsc{Graves-CPA} & \textbf{4780}                    & \textbf{237}                         & \textbf{1830}                     & \textbf{4972}               \\ \bottomrule
        \end{tabular}%
\end{table}

\subsubsection*{SV-Comp 2023 Evaluation}
As mentioned in Section~\ref{selectionRelated}, we entered a prototype of \textsc{Graves} in SV-Comp 2022 called \textsc{Graves-CPA}~\cite{GRAVES-SVCOMP22}. Inspired by the tool PeSCo, \textsc{Graves-CPA} selects from various configurations of the CPAChecker tool. Since we were unable to recreate WLJ's study to compare it against all of our metrics, we decided to perform a head-to-head comparison of PeSCo, which is based on the WLJ technique, and \textsc{Graves-CPA}. To make it a fair comparison, we have updated \textsc{Graves-CPA} to use the same CPAChecker configuration file as PeSCo, with the exception of the algorithm selector. We have also updated \textsc{Graves-CPA} to use our new GNN architecture.

Table~\ref{tab:SVCOMP23} shows the results of our evaluation of PeSCo and \textsc{Graves-CPA} on the SV-Comp 2023 Reach Safety benchmarks in four categories. We limit our evaluation to Reach Safety as PeSCo and \textsc{Graves-CPA} select from CPAChecker configurations which solve problems in Reach Safety categories. Results are reported using SV-Comp's scoring rules which weights the various subcategories in each category equally and greatly penalizes incorrect results~\cite{beyer2022progress}. 

In all four categories, \textsc{Graves-CPA} outperforms PeSCo. Reach Safety and Software Systems are standalone categories, while Falsification and Overall are meta-categories which are combine the scores of Reach Safety and Software Systems. PeSCo and \textsc{Graves-CPA} perform very similarly in Reach Safety. Software Systems problems deal with large, real world software like Linux device drivers, GNU coreutils, and AWS libraries. The fact the \textsc{Graves-CPA} has a score three times higher than PeSCo suggests that the \textsc{Graves} technique generalizes better to the kind of systems real developers deal with.

The Falsification category is meant to test how well tools can find violations in a specification. For this category, tools only receive scores when they report that they found a violation, no matter if this is correct or not. \textsc{Graves-CPA} greatly outperforms PeSCo in this category, outscoring it 40 to 1. Falsification techniques generate counterexamples to specifications, allowing developers to see concrete examples of bugs. Since it wins in both Reach Safety and Software Systems, it comes as no surprise that \textsc{Graves-CPA} beats PeSCo in the Overall category.

\noindent\Ovalbox{\begin{minipage}{\dimexpr\textwidth-2\fboxsep-2\fboxrule\relax}
    \textsc{Graves} outperforms all other algorithm selectors across every metric we evaluate. We find it selects the optimal verifier 60\% of the time, and selects a verifier which can correctly solve a verification instance 86\% of the time. In a simulated competition setting, \textsc{Graves} outperforms a competing algorithm selector in all categories they are evaluated in.
\end{minipage}}

\begin{figure}
    \includegraphics[width=\linewidth]{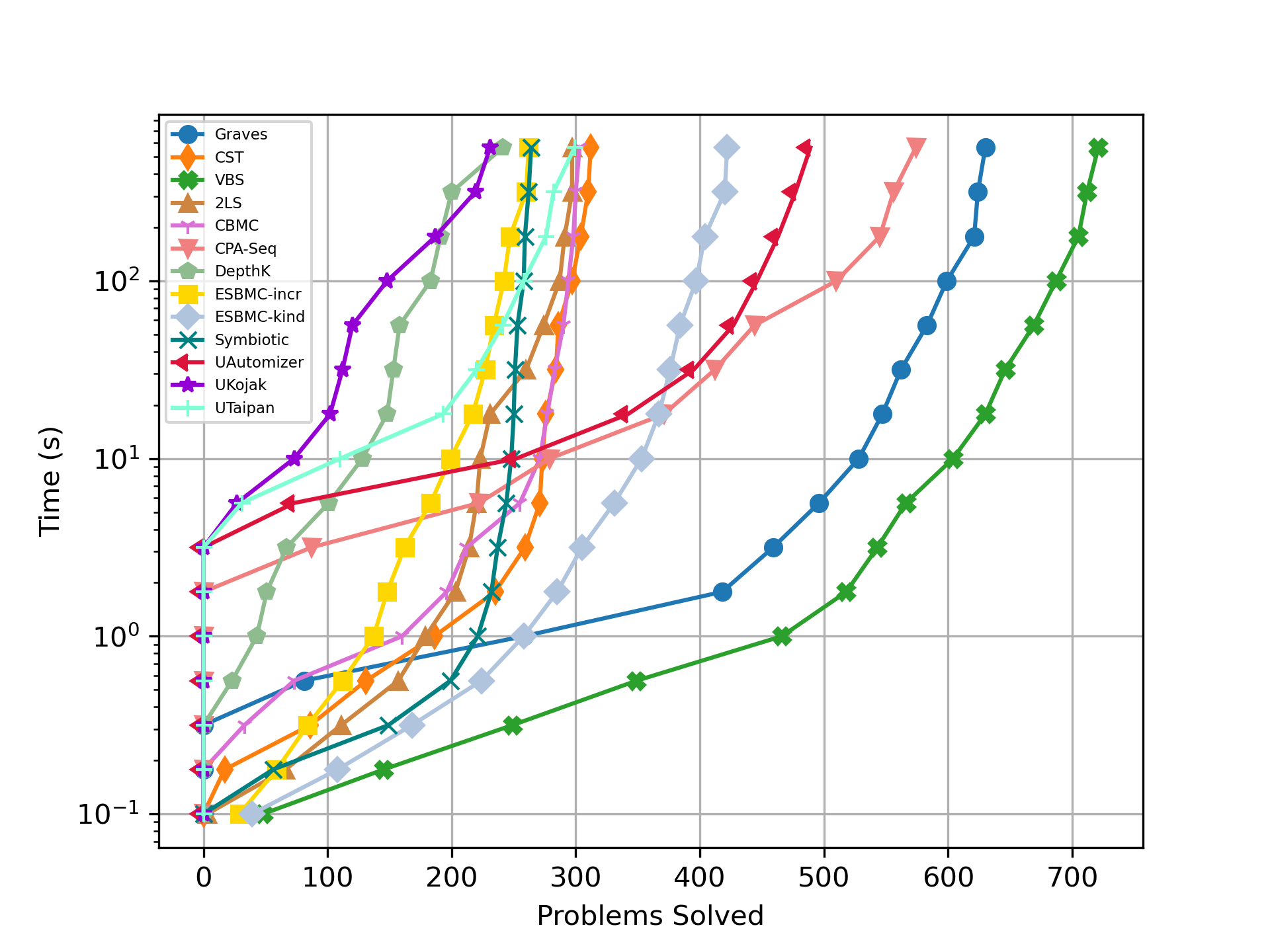}
    \captionof{figure}{Tool performance given a maximum runtime. For each tool (verifier or algorithm selector), we plot the maximum amount of problems the tool could solve given a maximum runtime. For example, given 10 seconds per problem, \textsc{Graves} can solve approximately 525 problems from the test set. For timeout values greater than 1s, \textsc{Graves}'s selected verifier is able to outperform any single verifier. Even with its overhead, \textsc{Graves} can select verifiers which allow it to verifier more problems than any single verifier.}
    \label{fig:cactusPlot}
\end{figure}

\subsection{Selection in Practice}
Across all three metrics, we find that \textsc{Graves} outperforms all selectors at the task of prediction. Up to this point, we have not taken into account \textsc{Graves}' overhead, which includes building a graph from the program, running the GNN on said graph, and forming a prediction. If this overhead is too high, it can outweigh the potential benefit of \textsc{Graves}' selection. To address this and \textbf{RQ2}, we compare \textsc{Graves}' selected verifier against each verifier in its portfolio, and include the cost of \textsc{Graves}'s overhead. 

Figure~\ref{fig:cactusPlot} shows, given a maximum runtime per problem in \changed{the test set, how many verification problems a given verifier or algorithm selector can solve, e.g. ESBMC-kind can solve almost 400 problems in the test set when given a timeout value of 100 seconds for each problem.} We list three algorithm selectors, \textsc{Graves}, CST, and the virtual best selector (VBS). VBS is a theoretical selector which makes optimal selections and has no overhead. We list it to show the upper bound on verification performance, given the test set and portfolio of verifiers. \changed{For the purposes of this study, we implement VBS as a lookup table which returns the time of the fastest verifier which could correctly verify the problem. \textsc{Graves} and CST include both the time it takes to make a selection as well as the time verification takes with the selected verifier. VBS only includes the verification time of the optimal verifier. Note that VBS does not solve all 833 problems in the test set. This is due to the fact that 106 of the problems in the test set cannot be solved by any verifier in the portfolio of verifiers.}

When the maximum runtime value is greater than 1s, \textsc{Graves} is able to verify more problems than any single solver or CST. \textsc{Graves} is able to select verifiers that outperform CPA-Seq, the single best verifier, when \textsc{Graves} and the verifier it selects are given a maximum runtime of 56s and CPA-Seq is given a maximum runtime of 900s. Given a maximum runtime of 900s, \textsc{Graves} is able to verify 76\% of the test set, while VBS can solve 87\%, meaning \textsc{Graves} is only 15\% away from optimal.

On the test set, \textsc{Graves}' average overhead is 1s. In the best case, it took \textsc{Graves} 0.3s to form a prediction. In the worst case, it took 25s. On average, \textsc{Graves} spent 46\% of the time forming a prediction. This may seem like \textsc{Graves} is wasting time that could be used to run a verifier, but selection allows \textsc{Graves} to verify 9.3\% more problems than the best single verifier and on average 184\% faster. Note that this implementation of Graves is a proof of concept and has not been optimized for performance. It stands to reason an implementation optimized for performance would decrease the selection overhead, increasing \textsc{Graves} performance.

\noindent\Ovalbox{\begin{minipage}{\dimexpr\textwidth-2\fboxsep-2\fboxrule\relax}
    \textsc{Graves} is able to verify the same amount of problems as the single best verifier in 6\% of the time, or 9.3\% more problems in the same amount of time.
\end{minipage}}

\subsection{Ablation Study}\label{RQ3}
To analyze the importance of the components of the network architecture and address \textbf{RQ3}, we perform two ablation studies. Looking at Figure~\ref{fig:gnn_architecture}, there are 3 components to the GNN in \textsc{Graves}: a series of EGC layers, a jumping knowledge layer, and a pooling layer. We first vary the number of EGC layers and whether or not their is a jumping knowledge layer. These are directly related as the jumping knowledge layer takes in the output of each EGC layer so the network can learn using each output. The second ablation study looks at the different pooling mechanisms the network uses.

\begin{figure}[]
    \centering
    \begin{minipage}[c]{0.49\textwidth}
      \centering
      \resizebox{\linewidth}{!}{
        \begin{tabular}{@{}llrr@{}}
            \toprule
            \multicolumn{1}{c}{\begin{tabular}[c]{@{}c@{}}EGC \\ Layers\end{tabular}} &
            \multicolumn{1}{c}{\begin{tabular}[c]{@{}c@{}}Jumping \\ Knowledge\end{tabular}} &
            \multicolumn{1}{c}{Success} &
            \multicolumn{1}{c}{Spearman} \\ \midrule
            0 & N/A   & 0.854 $\pm$ 0.006 & 0.706 $\pm$ 0.006 \\
            1 & False & 0.855 $\pm$ 0.014 & 0.728 $\pm$ 0.005 \\
            1 & True  & \textbf{0.863 $\pm$ 0.006} & \textbf{0.735 $\pm$ 0.005} \\
            2 & False & 0.856 $\pm$ 0.011 & 0.729 $\pm$ 0.008 \\
            2 & True  & 0.855 $\pm$ 0.008 & \textbf{0.735 $\pm$ 0.005} \\ \bottomrule
        \end{tabular}}
    \end{minipage}
    \begin{minipage}[c]{0.49\textwidth}
      \includegraphics[width=\textwidth]{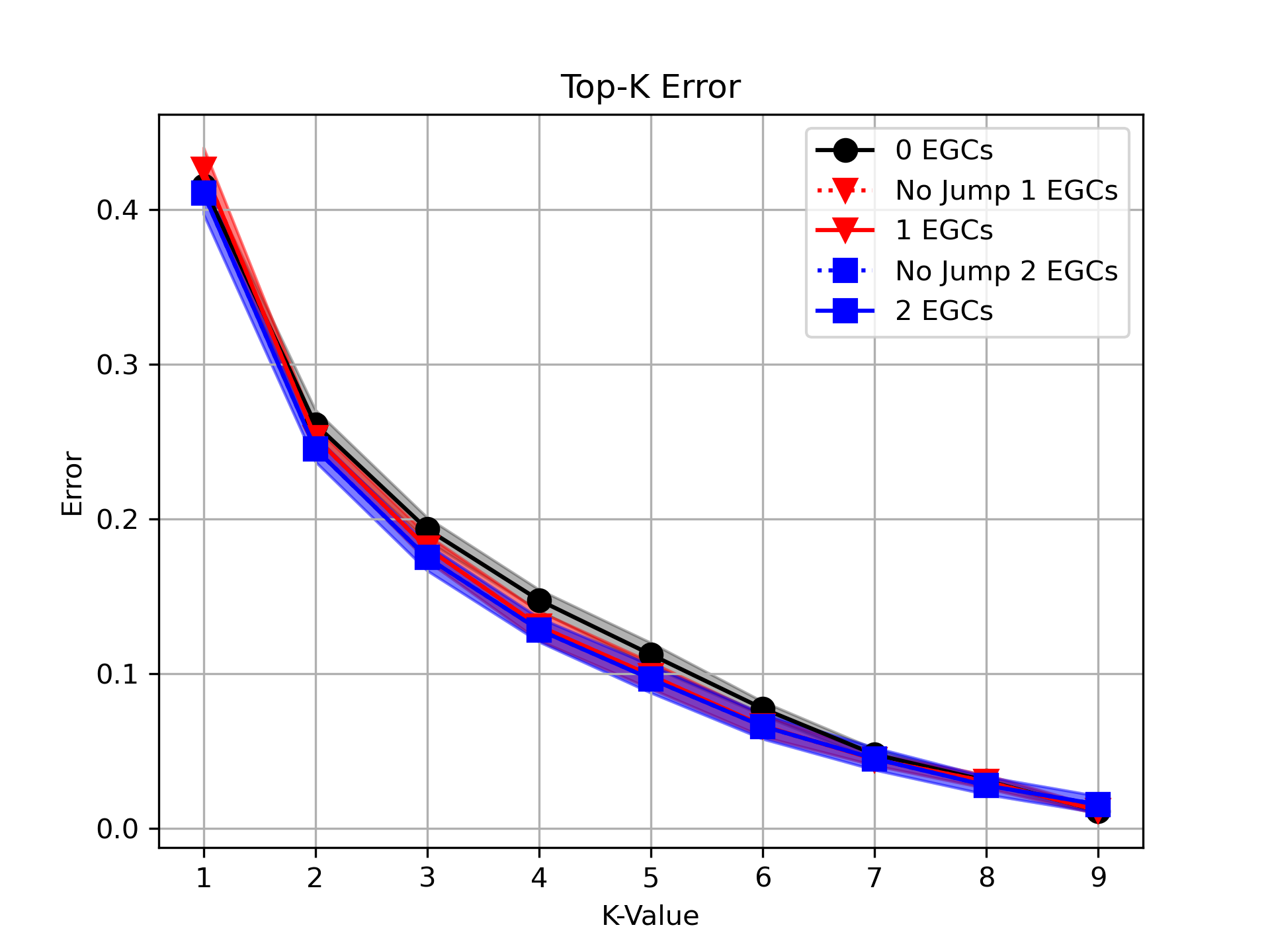}
    \end{minipage}
    \caption{Message Passing Layer Ablation Study Results. EGC layers improve \textsc{Graves}' prediction. The number of EGC layers is of less importance. The jumping knowledge layer has a slight, but noticeable effect on the accuracy.}
\label{figure:layer_ablation}
\end{figure}

\begin{figure}[]
    \centering
    \begin{minipage}[c]{0.49\textwidth}
      \centering
      \resizebox{\linewidth}{!}{
        \begin{tabular}{@{}lllrr@{}}
            \toprule
            \multicolumn{1}{c}{\begin{tabular}[c]{@{}c@{}}Max \\ Pooling\end{tabular}} &
              \multicolumn{1}{c}{\begin{tabular}[c]{@{}c@{}}Mean \\ Pooling\end{tabular}} &
              \multicolumn{1}{c}{\begin{tabular}[c]{@{}c@{}}Attention \\ Pool\end{tabular}} &
              \multicolumn{1}{c}{Success} &
              \multicolumn{1}{c}{Spearman} \\ \midrule
            True  & False & False & \textbf{0.855 $\pm$ 0.010} & 0.730 $\pm$ 0.006 \\
            False & True  & False & 0.840 $\pm$ 0.010 & 0.718 $\pm$ 0.007 \\
            False & False & True  & 0.843 $\pm$ 0.008 & 0.717 $\pm$ 0.006 \\
            True  & True  & False & 0.854 $\pm$ 0.007 & 0.733 $\pm$ 0.006 \\
            True  & False & True  & 0.853 $\pm$ 0.008 & \textbf{0.738 $\pm$ 0.005} \\
            False & True  & True  & 0.854 $\pm$ 0.010 & 0.726 $\pm$ 0.007 \\
            True  & True  & True  & \textbf{0.855 $\pm$ 0.008} & 0.735 $\pm$ 0.005 \\ \bottomrule
            \end{tabular}}
    \end{minipage}
    \begin{minipage}[c]{0.49\textwidth}
        \includegraphics[width=\linewidth]{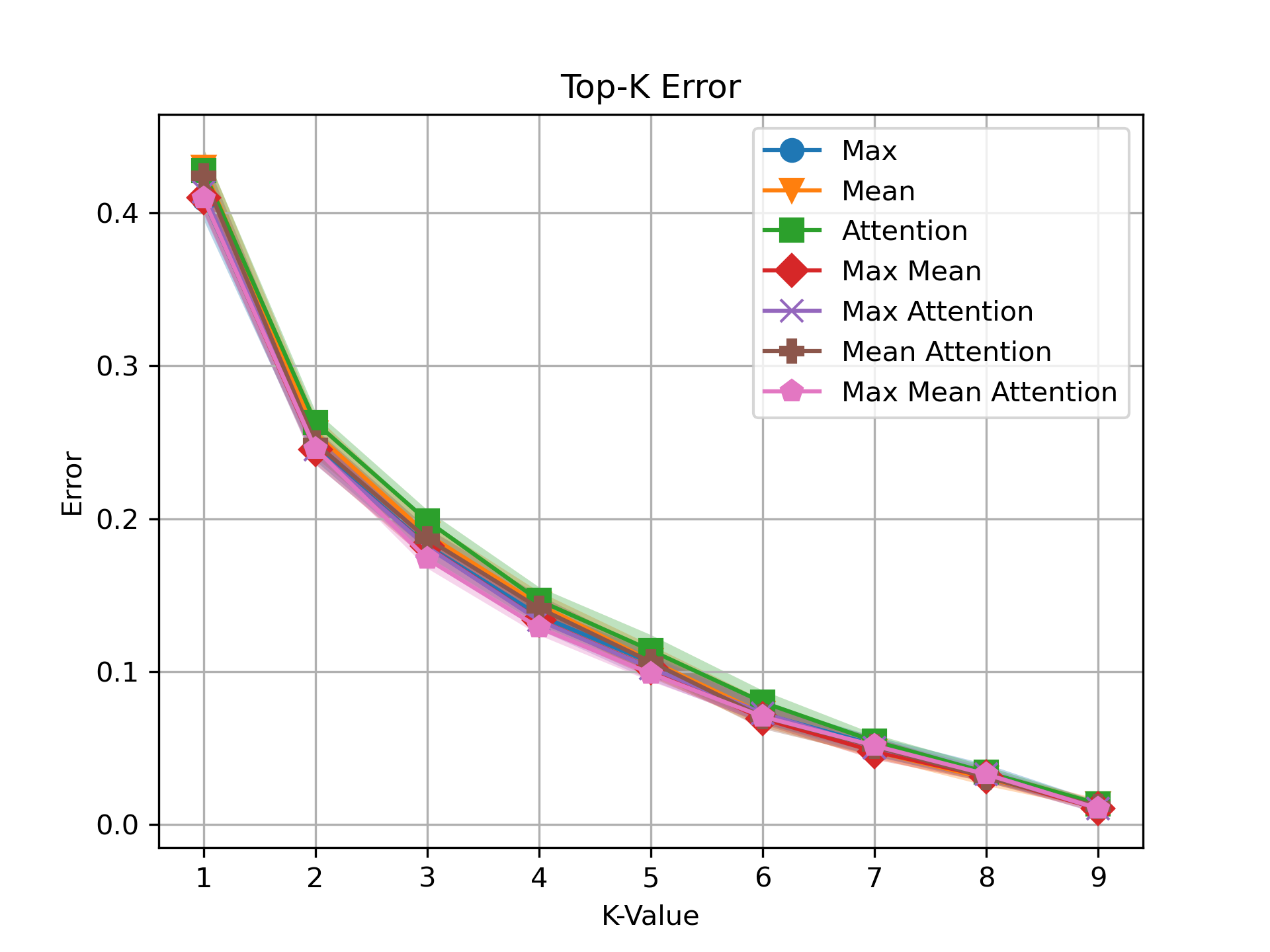}
    \end{minipage}
    \caption{Pool Ablation Study Results.}
\label{figure:pool_ablation}
\end{figure}
\subsubsection{Message Passing Ablation}
Figure~\ref{figure:layer_ablation} shows how \textsc{Graves}' networks perform across all three metrics when the number of EGC layers are varied from 0-2 and when jumping knowledge layers are present or absent. As a reminder, the architecture used in previous studies has two EGC layers and a jumping knowledge layer. We find that EGC layers improve performance roughly 1-4\% over having no EGC layers, depending on the metric. Networks with a single EGC layer have the best successful verifier accuracy performance. Networks with two EGC layers have the best performance for Top-K. Networks with 1 or 2 EGC layers perform equally on Spearman Rank Correlation. Note that networks with no EGC layers still outperform our baseline techniques. This shows that the pooling techniques and neural network are powerful on their own.

The jumping knowledge layer appears to play a small, but noticeable role in the networks abilities to form predictions. Each network with a jumping knowledge layer performs about 1\% better than their counterpart without a jumping knowledge layer, with the exception of the networks with two EGC layers when evaluated using the success metric. We also find the variance tends to decrease when a jumping knowledge layer is added.

\subsubsection{Pooling Layer Ablation}
Figure~\ref{figure:pool_ablation} shows how \textsc{Graves}' networks perform across all three metrics using all combinations of max, mean, and attention pools. Of the three pools, the max pool has the greatest effect on network performance. When only one pool is included, the network with the max pool performs 1.7\% better than either the mean or attention pools. The combination of a max and an attention pool performs best in Spearman Rank Correlation. Using all three pools, the network is able to achieve the best performance in successful verifier accuracy and Top-K for K<6. 

Max pooling layers have been shown to allow the GNN to learn to create representations that identify distinct features, such as an infinite loop or a nonlinear path condition~\cite{xu2018powerful}. If there are certain behaviors that cause different verifiers to struggle or succeed, the GNN may learn to preserve these characteristics allowing the prediction network to learn to differentiate them. The attention pool appears to be slightly more important than the mean pool, which indicates that the structure of the graph is more important than the distribution of the features~\cite{xu2018powerful}. This is reasonable as the number of loops, conditional statements, or integers is less important than how they interact when considering verification complexity.

\noindent\Ovalbox{\begin{minipage}{\dimexpr\textwidth-2\fboxsep-2\fboxrule\relax}
    EGC and jumping knowledge layers both improve \textsc{Graves}' network's ability to make predictions. All three pooling layers benefit \textsc{Graves}' network, with max pooling being the most important.
\end{minipage}}

\subsection{Threats to Validity}
In the following section, we discuss potential threats to the validity of our model evaluation experiments.

\subsubsection{Internal Threats}
A potential internal threat to this study is our implementation of the approach. To mitigate this, we inserted assertions in our implementation to ensure it matched our specifications. This included ensuring all edges only referenced nodes that exist in the AST, every verification problem in our dataset had a specification and a label, and that during training our model made valid predictions. We also performed sanity checks on the graphs, such as checking that our ASTs contained no cycles.

\subsubsection{External Threats}
An external threat to validity revolves around the training data. To produce our graphs, we used only C programs and Clang's AST tokens. Different compilers and languages may have different tokens, though compilation systems for standardized languages tend to have very similar internal representations. As a result, we do not know how \textsc{Graves} would transfer to different languages or ASTs. 

The 2018 SV-Comp dataset contains many realistic examples of C software, such as Linux device drivers and the GNU Coreutils. However, there are also many unrealistic examples, such as programs consisting of 20 SLOC containing a simple loop. These unrealistic examples may affect how our models generalize to examples in the wild.

With these issues in mind, we proceeded with this dataset for several reasons. For one, many examples, while small, could not be verified by all the verifiers used to evaluate \textsc{Graves}. This shows that these small portions of source code may give insight into what causes these tools to falter. Previous techniques also used the SV-Comp 2018 dataset. By using the same dataset, we were able to perform a direct comparison between techniques. Finally, and most importantly, in order to perform training, we need the ground truth for each verification problem and building a new dataset was outside the scope of this project. While this dataset is not at the scale of some machine learning datasets, we were able to obtain results that, for the most part, exceeded or were comparable to the previous state-of-the-art.
\section{Model Interpretability}\label{model_interpretability}
To address \textbf{RQ4}, we perform a qualitative study using the GNNExplainer~\cite{ying2019gnnexplainer} technique to examine which portions of the program have the greatest affect on \textsc{Graves}' predictions.
For each verification technique, we randomly select 10 programs where \textsc{Graves} correctly selects said technique as the best performing algorithm.  
We limit our selection of programs to those that produce graphs with less than 500 nodes.
Graphs above this size were deemed to be too costly to analyze by hand, even by an expert in program representations and verification techniques.

GNNExplainer produces a score $s_i\in (0,1)$ for each edge $e_i$ in a graph which has $n$ edges. The higher the score, the larger the affect the edge has on the networks ability to make its prediction. Since the values of $s_i$ can vary widely for a given graph, choosing a threshold to use across graphs was problematic. Instead, we highlight the 10 edges to which GNNExplainer gives the highest score.

To evaluate our graphs, we went through the process of open coding~\cite{corbin2014basics}. Open coding is a technique in qualitative data analysis which summarizes data using various terms or short phrases, called codes, identified by a group of researchers. These codes can then reveal different patterns in the data, such as reoccurring themes or outliers.
\subsection{Case Study}
\begin{figure}
    \centering
\begin{lstlisting}[language=C,basicstyle=\footnotesize]
int main(void) {
    unsigned int x = 0x0ffffff0;
    while (x > 0) {
        x -= 2;
    }
    __VERIFIER_assert(!(x % 2));
}   
\end{lstlisting}    
\caption{The program \texttt{simple\_true\_unreach-call4.i}, which uses SV-Comp's assertion syntax of ``\_\_VERIFIER\_assert(condition)'' as opposed to the built in ``assert(condition)''. From the verification techniques we observed, bounded model checking with K-Induction was fastest at proving that x is even at the end of main.}\label{fig:RQ4Code}
\end{figure}

\begin{figure}
    \centering
       \includegraphics[width=\textwidth]{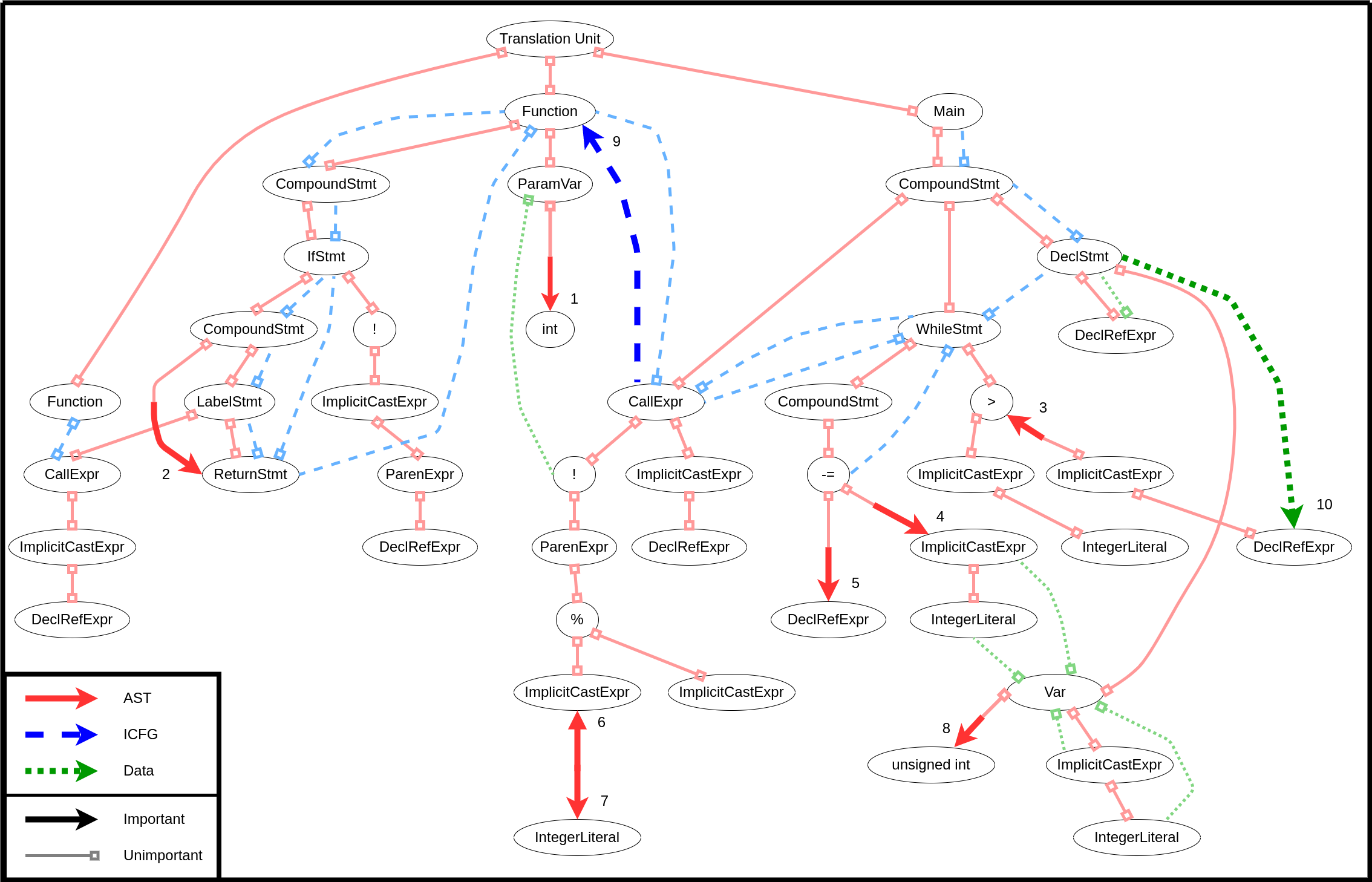}
    
    \caption{This is the program graph \textsc{Graves} generated of \texttt{simple\_true\_unreach-call4.i} from Figure~\ref{fig:RQ4Code}. The GNNExplainer determines the assertion condition (edges labeled 1, 6, and 7), the assertion call (edges 2 and 9), and variable referenced in both the assertion and loop condition (edges 3, 4, 5, 6, and 10) are important. The assertion condition and call are directly related to an error state being reached. Using K-Induction, the verifier can reason about the semantics of the loop without having to unwind it over 100,000,000 as a bounded model checker without K-Induction or a symbolic execution tool would.}\label{fig:RQ4Graph}
\end{figure}

To illustrate the coding process, we examine Figure~\ref{fig:RQ4Graph}. 
This is the graph representation of the program \texttt{simple\_true\_unreach-call4.i} shown in Figure~\ref{fig:RQ4Code}. 
The 10 highest scoring edges are emboldened, end in arrows, and are labeled. The remaining edges are less saturated and end in a square. AST edges are undirected, but GNN's operate on directed graphs. We represent them using a single, bidirectional edge. If one of the directed edges is important, we embolden half of the edge and arrow in the appropriate direction. If both directed edges are important, we embolden both directions and arrows.

We start by evaluating the AST edges of the graph. The edge labeled 1 goes from the parameter variable of the \texttt{\_\_VERIFIER\_assert} function to its type. 
This is the type of the assertion condition, so we label this edge ``Assertion Condition''. The edge labeled 2 goes to the return statement of the assertion function and is labeled ``Assertion Function Return''. The third edge goes to the less than symbol in the while loop's condition. Since the while loop will iterate over 1000 times, we label this edge ``Loop Condition (Large Bound)''. The fourth and fifth edge both deal with decrementing the variable \texttt{x}, the former referencing \texttt{x} itself and the latter referencing the value being subtracted. The variable \texttt{x} is referenced both in the loop condition and assertion condition. Variables referenced in assertion conditions are called ``Directly Dependent Variables'', as the assertion's satisfiability is directly dependent on their value. There are also ``Indirectly Dependent (X) Variables'', which are variables which affect the value of directly dependent variables. The ``X'' value is an integer which determines how far down the dependence chain the variable is. Thus, both edges 4 and 5 are labeled ``Update of Loop Condition Variable (Large Bound)'' and ``Update of Directly Dependent Variable''. Edges 6 and 7 are directed to and from the integer literal in the assertion condition. Like edge 2, we label these ``Assertion Condition''. Finally, edge 8 is directed at the type declaration of \texttt{x}. Like edges 4 and 5, we label this edge both ``Update of Loop Condition Variable (Large Bound)'' and ``Update of Directly Dependent Variable'' as its initial value is determined in its declaration.

Next, we label the ICFG and Data edges. In the case of this graph, there is only one of each: edges 9 and 10. Edge 9 is the call edge from the assertion call to the function, so we label it ``Assertion Call''. Edge 10 is a def-use edge from the definition of \texttt{x} to its use in the loop condition. Since \texttt{x} is both a loop condition variable and a directly dependent variable, we label edge 10 both ``Use of Directly Dependent Variable'' and ``Use of Loop Condition Variable''.

\subsection{Results}

\begin{table}[]
    \centering
    \caption{Code category breakdown per verification technique. For each category, we list the the rate at which codes fall into said category, e.g., 19.2\% of BMC codes deal with the reachability of an error state. Across all categories, codes in the assertion satisfiability category are most used. Each verification technique is attempting to determine if an error state is reachable, which occurs when a assertion can be violated.}
    \begin{tabular}{@{}lrrrr@{}}
        \toprule
        \multicolumn{1}{c}{Category} & \multicolumn{1}{c}{BMC} & \multicolumn{1}{c}{BMC+K} & \multicolumn{1}{c}{CEGAR} & \multicolumn{1}{c}{SymEx} \\ \midrule
        Assertion Satisfiability     & 0.538                   & 0.582                     & 0.409                     & 0.373                     \\
        Branching (Excluding Loops)  & 0.029                   & 0.064                     & 0.082                     & 0.109                     \\
        Error State Reachability       & 0.192                   & 0.109                     & 0.173                     & 0.273                     \\
        Looping (Small Bound)        & 0.135                   & 0.055                     & 0.064                     & 0.155                     \\
        Looping (Large Bound)        & 0.077                   & 0.173                     & 0.227                     & 0.064                     \\
        Miscellaneous                        & 0.029                   & 0.018                     & 0.045                     & 0.027                     \\ \bottomrule
        \end{tabular}
    \label{tab:codesOccurences}
\end{table}

\begin{table}[]
    \centering
    \caption{Two most common codes per category. To give examples of the codes in each categories, we list the two most common codes per category. A full list of codes, their rate of occurrence, and their category can be found in the appendix.}
    \resizebox{\columnwidth}{!}{%
    \begin{tabular}{@{}lrrrrl@{}}
    \toprule
    \multicolumn{1}{c}{Code} & \multicolumn{1}{c}{BMC} & \multicolumn{1}{c}{BMC+K} & \multicolumn{1}{c}{SymEx} & \multicolumn{1}{c}{CEGAR} & \multicolumn{1}{c}{Category}  \\ \midrule
    Assert Condition             & 0.115 & 0.182 & 0.155 & 0.118 & Assertion Satisfiability \\
    Branch Condition             & 0.019 & 0.027 & 0.100 & 0.064 & Branching (No Loops)     \\
    Conditional Statement        & 0.010 & 0.036 & 0.009 & 0.018 & Branching (No Loops)     \\
    Directly Dependent Function Return              & 0.096 & 0.000 & 0.100 & 0.082 & Error Reachability       \\
    Dead Code                    & 0.019 & 0.018 & 0.009 & 0.018 & Misc.                    \\
    Error Call                   & 0.019 & 0.055 & 0.045 & 0.027 & Error Reachability       \\
    Input Fun                    & 0.000 & 0.000 & 0.018 & 0.018 & Misc.                    \\
    Loop Condition (Large Bound) & 0.010 & 0.027 & 0.000 & 0.045 & Loop (Large Bound)       \\
    Loop Condition (Small Bound) & 0.038 & 0.027 & 0.027 & 0.000 & Loop (Small Bound)       \\
    Update Of Directly Dependent Var                & 0.087 & 0.182 & 0.136 & 0.073 & Assertion Satisfiability \\
    Update Of Loop Condition Var (Large Bound)      & 0.019 & 0.064 & 0.000 & 0.009 & Loop (Large Bound)       \\
    Update Of Recursion Condition Var (Small Bound) & 0.000 & 0.000 & 0.055 & 0.036 & Loop (Small Bound)       \\ \bottomrule
    \end{tabular}%
    }\label{tab:codes}
\end{table}

Of the 40 programs selected, we identified 64 codes for the 400 edges. Several edges could be described with two codes, which left us with 434 uses of our codes. Each code can be grouped into one of six categories: assertion satisfiability, branching (excluding loops), error reachability, looping (small bound), looping (large bound), and miscellaneous.
We define small bounds to be less than 1000 and large bounds to be greater than 1000, including infinite loops and loops bounded by user input. The miscellaneous category includes codes which do not fit into any other category. 

Table~\ref{tab:codesOccurences} summarizes the results of open coding. Each cell displays what portion of codes for a given technique belong to a given category. For example, 53.8\% of BMC edges can be categorized as relating to assertion satisfiability. Table~\ref{tab:codes} shows the two most common codes for each category. A full list of the codes can be found in the appendix.

\subsubsection{Overarching Patterns}\label{RQ4Overall}
Overall, the codes imply that the network is learning the reachability problem. To solve reachability problems, verification tools must determine whether or not an error state, such as an assertion violation, is reachable. For all techniques, the combination of the assertion satisfiability and error state reachability categories account for the majority of codes. The remaining categories, excluding Miscellaneous, relate to the intricacies of each verification technique.

\changed{
    AST edges make up 61.2\% of edges in the 40 program graphs. ICFG edges make up 20.0\% of the edges, and data dependence edges make up the remaining 18.8\%. These proportions stay roughly the same when looking at the edges the GNNExplainer gives the highest score. 59.5\% of the top 10 highest scoring edges from all 40 program graphs are AST edges. 19.3\% are ICFG edges, and 21.2\% are data dependence edges. Overall, this suggests each edge set is significant to the GNN's ability to make predictions as there is no large disparity. Data dependence edges appear to be slightly more important than the other two sets as it is the only edge set that has a greater proportion of edges in the top 10 highest scoring edges than its overall proportion.
}

\subsubsection{Bounded Model Checking} \label{RQ4BMC}
When an error state is reachable, BMC tools must be able to prove this within a given loop unrolling bound---usually provided as a configuration parameter. When an error state is unreachable, all looping or recursive behavior should have a finite, generally small, number of iterations.

Besides the edges dealing with the reachability problem, the GNNExplainer identifies Looping (Small Bound) to be the most important category for BMC. In particular, 85\% of the edges in the Looping (Small Bound) category for BMC are either variables reference in a loop condition or iterators. These variables decide how many iterations of the loop are executed, which is important as BMC struggles with large bounds. 

There are two programs where an edge is categorized as Looping (Large Bound). Neither of these programs can be solved by any of the algorithms. Typically, when none of the techniques can solve a problem, the model predicts BMC. The model is most likely identifying patterns all algorithms struggle with, and then defaults to BMC since it tends to report unknown first. BMC unrolls loops to a certain bound and then attempts to prove the property. If it can't prove the property or find a counterexample within the bound, it will report unknown, whereas SymEx will continue to search the state space, BMC+K will increase K and try again, and CEGAR will continuously refine it abstraction. 

\subsubsection{Bounded Model Checking with K-Induction} \label{RQ4BMC+K}
K-Induction is a generalization of the induction principle. While induction attempts to prove both the base case and the n+1 case, K-Induction attempts to prove both the base case and the n+k case. By integrating K-Induction into BMC, tools do not have to exhaustively explore looping and recursive behavior. Instead, they can attempt to prove the base case and the n+k case are safe.

Nearly a quarter of all edges deemed important by the model for BMC+K graphs deal with looping behavior, whether it be the loop itself or a loop index variable which bounds the loop's execution. 75\% of these codes deal with loops with large bounds. If there is a small bound, K-Induction may be more expensive than tools which unwind the loop to completion and verify the program. This may explain why the network looks for codes dealing with large bounds on loops, as the loop needs to have a high enough bound for K-Induction to be worth its cost.

\subsubsection{Counter-example Guided Abstraction Refinement}\label{RQ4CEGAR}
CEGAR based tools begin verification with a relatively simple abstract domain and attempt to prove the specification. If the domain is too general to capture the branching pattern of the program, it may generate false positives which force the algorithm to refine the domain until it can either prove the property, or find a true counterexample to the property~\cite{clarke2000counterexample}.

As stated in Section~\ref{RQ4Overall}, CEGAR problems identify edges dealing with the reachability problem as important. Proving reachability and satisfiability are related to the domain of the CEGAR tool. The domain must be specific enough to prove the existence or absence of a counterexample. If a simple domain suffices, the CEGAR tool can quickly verify the program. If there is a complex branching structure with nonlinear constraints, the tool will go through many refinements before it can prove the property.
Similar to the BMC+K graphs, CEGAR graphs identify edges dealing with loops with large bounds. CEGAR tools can reasons about loops of arbitrary sizes to find a fixed point in relation to their abstract domain. 

\subsubsection{Symbolic Execution} \label{RQ4SymEx}
Symbolic executors operate by exploring the execution tree of a program. While exploring a path in the tree, if they encounter an error state, they check if an assignment to free variables (inputs) allows for the path to said state to be feasible. As a result, they struggle with large or unbounded loops and recursion, but excel with programs with complex branching structures and loops with small bounds.

Like BMC, the programs where symbolic execution was selected contained far more codes dealing with loops with a small bound. They also contained the most problems dealing with error state reachability and branching. If the conditions leading to an assertion are infeasible, the symbolic execution can cease exploring paths on that branch of the execution tree. When there are large, potentially unbounded loops, the model also identifies assumption statements which place restrictions on the variables that bound the loop.

\noindent\Ovalbox{\begin{minipage}{\dimexpr\textwidth-2\fboxsep-2\fboxrule\relax}
    \textsc{Graves'} model identifies edges relating to the error reachability problem. For each algorithm, \textsc{Graves}' model identifies portions of the program that an expert would use to predict the given algorithm's success.
\end{minipage}}

\subsection{Threats to Validity}
In the following section, we describe potential threats to the validity of our experiment.
\subsubsection{Internal Threats}
Due to the fact that this study is a qualitative study, potential internal threats to validity come from the bias of the researchers. The codes we used come from terms and definitions from the community at large~\cite{ferrante1987program,ottenstein1984program,aho2007compilers}. Thus, any code, edge pair should be determinable by other experts. 

\subsubsection{External Threats}
As with the experiments in Section~\ref{model_evaluation}, a potential threat to this study is the choice of data. In order to be able to reason about the graph as a whole, we limited the size of graphs we evaluated. This was necessary to make the problem more tractable. We randomly selected the programs for this study from this abbreviated set in order to avoid bias.


\section{Conclusion} \label{conclusion}
In this work, we have proposed \textsc{Graves}, a technique to perform algorithm selection on program verifiers using graph neural networks. \textsc{Graves} automatically generates a graph representation of a program using traditional program graph representations that preserve semantic and syntactic components of said program. 
Using graph neural networks, \textsc{Graves} learns to form graph feature vectors which encodes the structure of the graph into a fixed size vector.
\textsc{Graves} passes this vector to a simple feedforward neural network which scores verifiers on how likely it is that they could successfully verify the given program.

We evaluated \textsc{Graves} using three metrics on over 8000 programs against several baseline techniques. We found that \textsc{Graves} is superior to several state-of-the-art baseline techniques we evaluated on the problem of selecting a verifier for a given program and property by over 12\%. Further, \textsc{Graves} was able to verify 9\% more programs from our test set than any single verifier.

We performed a study to interpret how \textsc{Graves} determines which verification technique to select. To do this we looked at three fundamental techniques and one technique variant: CEGAR, symbolic execution, bounded model checking, and bounded model with K-Induction. We found that \textsc{Graves} was able to identify portions of the graph related to the verification problem at hand. We also found that it selected portions of the graph specific to the given algorithms approach.

Moving forward, we would like to explore other applications for GNNs and software engineering problems. While we only explored the problem of verification algorithm selection, there is reason to believe that this approach could produces strong results in the space of software engineering. We would like to explore improving our graphs using compiler optimizations, such as dead code elimination or loop unrolling, and approaching other software engineering problems, like test generation and fuzzing.

\section*{Acknowledgment}
We would like to thank Hongning Wang for his advice on graph neural networks and prediction systems. This material is based in part upon work supported by the U.S. Army Research Office under grant number W911NF-19-1-0054, by the DARPA ARCOS program under contract FA8750-20-C-0507, and by the U.S. Air Force Office of Scientific Research under award FA9550-21-1-0164.

\bibliographystyle{acm}
\bibliography{sections/references.bib}

\clearpage

\appendix
\section{Category Specific Training}\label{CategorySpecific}
\begin{table}[h]
\caption{Spearman Rank Correlation Results for Category specific training. When training selectors for specific specifications, WLJ is superior in all but the memory safety case. It may appear that \textsc{Graves} could benefit from category specific training, but Table~\ref{table:spearmanCategoryAllTrained} shows otherwise.}
\resizebox{\textwidth}{!}{
\begin{tabular}{lrrrrr}
\toprule
\textbf{Algorithm Selector} & \multicolumn{1}{c}{\textbf{Reach Safety}} & \multicolumn{1}{c}{\textbf{Termination}} & \multicolumn{1}{c}{\textbf{Memory Safety}} & \multicolumn{1}{c}{\textbf{Overflow}} \\ \midrule
\textsc{Graves}$_\text{default}$ & 0.696 $\pm$ 0.004 & 0.866 $\pm$ 0.005 & 0.765 $\pm$ 0.010 & 0.571 $\pm$ 0.021 \\\arrayrulecolor{lightgray}\hline
\textsc{Graves}$_\text{reach}$ & 0.712 $\pm$ 0.010 & 0.846 $\pm$ 0.007 & 0.762 $\pm$ 0.012 & 0.597 $\pm$ 0.005  \\\hline
\textsc{Graves}$_\text{term}$ & 0.695 $\pm$ 0.009 & 0.871 $\pm$ 0.007 & 0.761 $\pm$ 0.013 & 0.583 $\pm$ 0.011 \\ \hline
\textsc{Graves}$_\text{mem}$ & 0.697 $\pm$ 0.005 & 0.842 $\pm$ 0.010 & \textbf{0.767 $\pm$ 0.009} & 0.597 $\pm$ 0.005 \\ \hline
\textsc{Graves}$_\text{flow}$ & 0.680 $\pm$ 0.011 & 0.864 $\pm$ 0.008 & 0.765 $\pm$ 0.009 & 0.604 $\pm$ 0.019 \\\arrayrulecolor{black}\hline
WLJ$_{\text{reach}}$ & \textbf{0.719 $\pm$ 0.019} & 0.879 $\pm$ 0.021 & 0.647 $\pm$ 0.057 & 0.777 $\pm$ 0.046 \\ \arrayrulecolor{lightgray}\hline
WLJ$_{\text{term}}$ & 0.717 $\pm$ 0.020 & \textbf{0.881 $\pm$ 0.020} & 0.644 $\pm$ 0.054 & \textbf{0.779 $\pm$ 0.044} \\ \hline
WLJ$_{\text{mem}}$ & 0.715 $\pm$ 0.021 & 0.877 $\pm$ 0.019 & 0.649 $\pm$ 0.054 & 0.769 $\pm$ 0.042 \\ \hline
WLJ$_{\text{flow}}$ & 0.717 $\pm$ 0.020 & \textbf{0.881 $\pm$ 0.020} & 0.644 $\pm$ 0.054 & \textbf{0.779 $\pm$ 0.044} \\ \arrayrulecolor{black}\hline
CST & 0.180 $\pm$ 0.000 & 0.112 $\pm$ 0.000 & 0.175 $\pm$ 0.000 & 0.214 $\pm$ 0.000 \\ \arrayrulecolor{black}\hline
ISS & 0.309 $\pm$ 0.000 & 0.369 $\pm$ 0.000 & 0.470 $\pm$ 0.000 & 0.433 $\pm$ 0.000 \\ \arrayrulecolor{black}\hline
Random & 0.000 $\pm$ 0.016 & 0.001 $\pm$ 0.021 & 0.003 $\pm$ 0.012 & -0.004 $\pm$ 0.055 \\ \arrayrulecolor{black}\bottomrule
\end{tabular}}
\label{table:spearmanCategory}
\end{table}
Each column in Table~\ref{table:spearmanCategory} shows the Spearman correlation for a given technique and configuration when trained and evaluated on a specific problem type, denoted by the column header. We report the average and standard deviation for 10 selectors of each technique and configuration. 

We list five versions of \textsc{Graves}: default, reach, term, mem, and flow. Each of these use the GNN configuration we found to be optimal for the problem it derives its name from. 
The default configuration is the same configuration as used previously, 2 ECG layers and AST, ICFG, and Data edges in the graphs. 
The reach configuration includes 2 ECG layers and ICFG and Data edges in the graph. 
The term configuration includes 1 ECG layers and AST, ICFG, and Data edges in the graph.
The mem configuration includes 2 ECG layers and ICFG edges in the graph.
The flow configuration includes 4 EGC layers and AST and ICFG layers in the graph.

Similarly, we list 4 configurations of WLJ, each being optimal for the problem it derives its name from. Once again, these results come from the authors' evaluation of WLJ in \cite{richter2020algorithm}. Each configuration has an (i, j) pair as follows: reach=$(2,5)$, term=$(2,4)$, mem=$(1,5)$, flow=$(2,4)$.


In \cite{richter2020algorithm}, Richter et al. showed that training algorithm selectors for verification on specific problem types can provide significant gains in Spearman correlation. 
This may be due to their technique's machine learning component. As stated previously, there is not a convenient way to provide information about $\phi$ to their SVM's kernel. 
Thus, it cannot make decisions informed by the problem type. 
This is an issue as certain verifiers may perform well at proving one property properties, but suffer at proving another. 
Because of this, they stand to gain a lot from category specific training. 

\begin{table}[t]
\caption{Spearman Rank Correlation Results when the GNN is trained on all data. \textsc{Graves} only has marginal gains in the Overflow category when we do category specific training. In reach safety, we receive the same results, and in termination and memory safety, there is a noticeable improvement from training on all of the training data.}
\resizebox{\textwidth}{!}{
\begin{tabular}{@{}lrrrr@{}}
\toprule
\multicolumn{1}{c}{\textbf{Algorithm Selector}} & \multicolumn{1}{c}{\textbf{Reach Safety}} & \multicolumn{1}{c}{\textbf{Termination}} & \multicolumn{1}{c}{\textbf{Memory Safety}} & \multicolumn{1}{c}{\textbf{Overflow}} \\ \midrule
\textsc{Graves}$_{\text{full}}$ & 0.698 $\pm$ 0.005 & \textbf{0.882 $\pm$ 0.008} & 0.740 $\pm$ 0.019 & 0.593 $\pm$ 0.039 \\ \arrayrulecolor{lightgray}\hline
\textsc{Graves}$_{\text{reach}}$ & \textbf{0.711 $\pm$ 0.005} & 0.878 $\pm$ 0.013 & 0.733 $\pm$ 0.021 & 0.589 $\pm$ 0.050 \\ \hline
\textsc{Graves}$_{\text{term}}$ & 0.698 $\pm$ 0.005 & \textbf{0.882 $\pm$ 0.008} & 0.736 $\pm$ 0.019 & 0.593 $\pm$ 0.039 \\ \hline
\textsc{Graves}$_{\text{mem}}$ & 0.679 $\pm$ 0.004 & 0.871 $\pm$ 0.010 & \textbf{0.750 $\pm$ 0.008} & 0.584 $\pm$ 0.037 \\ \hline
\textsc{Graves}$_{\text{flow}}$ & 0.698 $\pm$ 0.012 & 0.879 $\pm$ 0.010 & 0.731 $\pm$ 0.013 & \textbf{0.617 $\pm$ 0.024} \\ \arrayrulecolor{black}\bottomrule
\end{tabular}}
\label{table:spearmanCategoryAllTrained}
\end{table}

Each column of Table~\ref{table:spearmanCategoryAllTrained} lists the results of \textsc{Graves} trained on the entire training set, but evaluated only on the tests set of one $\phi$ value. 
Once again, we show the results for the default networks and the optimal configurations for each category. The default configuration once again uses 2 EGC layers and AST, ICFG, and Data edges. 
The reach configuration includes 4 EGC layers and ICFG edges in the graph. 
The term configuration is the same as the default configuration.
The mem configurations includes 2 EGC layers and AST edges in the graph.
The flow configuration includes 1 EGC layers and ICFG edges in the graph. This chart shows that \textsc{Graves} can actually gain from training on all problem types. In fact, it can now outperform WLJ on termination.
\clearpage
\section{Full Open Coding Results}\label{appendix_data}
\begin{table}[h]
    \caption{Full list of Codes for Model Interpretability study. Each row lists a code, the rate each code occurs for each verification technique, and the category we place the code in.}
    \resizebox{!}{0.35\paperheight}{%
    \begin{tabular}{@{}lrrrrl@{}}
        \toprule
        \multicolumn{1}{c}{Code} & \multicolumn{1}{c}{BMC} & \multicolumn{1}{c}{BMC+K} & \multicolumn{1}{c}{SymEx} & \multicolumn{1}{c}{CEGAR} & \multicolumn{1}{c}{Category} \\ \midrule
        Assert Call                                           & 0.010 & 0.018 & 0.036 & 0.009 & Error Reachability       \\
        Assert Condition                                      & 0.115 & 0.182 & 0.155 & 0.118 & Assertion Satisfiability \\
        Assert Function                                       & 0.019 & 0.000 & 0.009 & 0.000 & Error Reachability       \\
        Assert Return                                         & 0.010 & 0.009 & 0.036 & 0.009 & Error Reachability       \\
        Assume Condition                                      & 0.019 & 0.045 & 0.000 & 0.027 & Assertion Satisfiability \\
        Branch Condition                                      & 0.019 & 0.027 & 0.100 & 0.064 & Branching (No Loops)     \\
        Call Of Indirectly Dependent (1) Function             & 0.000 & 0.000 & 0.000 & 0.009 & Error Reachability       \\
        Call Of Indirectly Dependent (2) Function             & 0.000 & 0.009 & 0.000 & 0.000 & Error Reachability       \\
        Conditional Statement                                 & 0.010 & 0.036 & 0.009 & 0.018 & Branching (No Loops)     \\
        Directly Dependent Function                           & 0.000 & 0.009 & 0.018 & 0.018 & Error Reachability       \\
        Directly Dependent Function Call                      & 0.000 & 0.009 & 0.000 & 0.000 & Error Reachability       \\
        Directly Dependent Function Return                    & 0.096 & 0.000 & 0.100 & 0.082 & Error Reachability       \\
        Dead Code                                             & 0.019 & 0.018 & 0.009 & 0.018 & Miscellaneous            \\
        Decl Of Directly Dependent Variable                   & 0.058 & 0.018 & 0.009 & 0.045 & Assertion Satisfiability \\
        Decl Of Indirectly Dependent (1) Variable             & 0.019 & 0.009 & 0.009 & 0.009 & Assertion Satisfiability \\
        Decl Of Indirectly Dependent (2) Variable             & 0.010 & 0.000 & 0.009 & 0.000 & Assertion Satisfiability \\
        Decl Of Loop Condition Variable (Large Bound)         & 0.010 & 0.000 & 0.000 & 0.000 & Loop (Large Bound)       \\
        Error Call                                            & 0.019 & 0.055 & 0.045 & 0.027 & Error Reachability       \\
        Error Function                                        & 0.019 & 0.000 & 0.000 & 0.000 & Error Reachability       \\
        Error Return                                          & 0.000 & 0.000 & 0.027 & 0.000 & Error Reachability       \\
        Indirectly Dependent (1) Function                     & 0.000 & 0.000 & 0.000 & 0.009 & Error Reachability       \\
        Indirectly Dependent (1) Function Return              & 0.010 & 0.000 & 0.000 & 0.009 & Error Reachability       \\
        Indirectly Dependent (1) Variable Update              & 0.000 & 0.000 & 0.000 & 0.027 & Assertion Satisfiability \\
        Input Function                                        & 0.000 & 0.000 & 0.018 & 0.018 & Miscellaneous            \\
        Loop Body (Small Bound)                               & 0.000 & 0.018 & 0.000 & 0.000 & Loop (Small Bound)       \\
        Loop Condition (Large Bound)                          & 0.010 & 0.027 & 0.000 & 0.045 & Loop (Large Bound)       \\
        Loop Condition (Nondet Bound)                         & 0.010 & 0.009 & 0.000 & 0.009 & Loop (Large Bound)       \\
        Loop Condition (Small Bound)                          & 0.038 & 0.027 & 0.027 & 0.000 & Loop (Small Bound)       \\
        Loop Header (Large Bound)                             & 0.010 & 0.009 & 0.000 & 0.000 & Loop (Large Bound)       \\
        Loop Header (Small Bound)                             & 0.010 & 0.000 & 0.009 & 0.000 & Loop (Small Bound)       \\
        Recursion Call (Nondet Bound)                         & 0.000 & 0.000 & 0.000 & 0.018 & Loop (Large Bound)       \\
        Recursion Call (Small Bound)                          & 0.000 & 0.000 & 0.027 & 0.009 & Loop (Small Bound)       \\
        Recursion Condition (Nondet Bound)                    & 0.000 & 0.000 & 0.000 & 0.009 & Loop (Large Bound)       \\
        Recursion Condition (Small Bound)                     & 0.000 & 0.000 & 0.018 & 0.009 & Loop (Small Bound)       \\
        Recursion Function (Small Bound)                      & 0.000 & 0.000 & 0.000 & 0.009 & Loop (Small Bound)       \\
        Recursion Function Decl (Nondet Bound)                & 0.000 & 0.000 & 0.000 & 0.009 & Loop (Large Bound)       \\
        Recursion Function Return (Small Bound)               & 0.000 & 0.000 & 0.018 & 0.000 & Loop (Small Bound)       \\
        Return Indirectly Dependent (1) Function              & 0.010 & 0.000 & 0.000 & 0.000 & Error Reachability       \\
        Return Of Recursion Function (Nondet Bound)           & 0.000 & 0.000 & 0.000 & 0.027 & Loop (Large Bound)       \\
        Type Def                                              & 0.010 & 0.000 & 0.000 & 0.009 & Miscellaneous            \\
        Type Of Directly Dependent Variable                   & 0.010 & 0.000 & 0.000 & 0.000 & Assertion Satisfiability \\
        Type Of Indirectly Dependent (1) Variable             & 0.000 & 0.000 & 0.000 & 0.009 & Assertion Satisfiability \\
        Update Loop Condition Variable (Infinite Loop)        & 0.000 & 0.000 & 0.064 & 0.009 & Loop (Large Bound)       \\
        Update Loop Condition Variable (Nondet Bound)         & 0.019 & 0.000 & 0.000 & 0.009 & Loop (Large Bound)       \\
        Update Of Directly Dependent Variable                 & 0.087 & 0.182 & 0.136 & 0.091 & Assertion Satisfiability \\
        Update Of Indirectly Dependent (1) Variable           & 0.135 & 0.064 & 0.045 & 0.018 & Assertion Satisfiability \\
        Update Of Indirectly Dependent (2) Variable           & 0.048 & 0.045 & 0.009 & 0.000 & Assertion Satisfiability \\
        Update Of Indirectly Dependent (3) Variable           & 0.000 & 0.027 & 0.000 & 0.000 & Assertion Satisfiability \\
        Update Of Loop Condition Variable (Infinite Loop)     & 0.000 & 0.000 & 0.000 & 0.009 & Loop (Large Bound)       \\
        Update Of Loop Condition Variable (Large Bound)       & 0.019 & 0.064 & 0.000 & 0.009 & Loop (Large Bound)       \\
        Update Of Loop Condition Variable (Nondet Bound)      & 0.000 & 0.055 & 0.000 & 0.009 & Loop (Large Bound)       \\
        Update Of Loop Condition Variable (Small Bound)       & 0.038 & 0.009 & 0.000 & 0.000 & Loop (Small Bound)       \\
        Update Of Recursion Condition Variable (Nondet Bound) & 0.000 & 0.000 & 0.000 & 0.055 & Loop (Large Bound)       \\
        Update Of Recursion Condition Variable (Small Bound)  & 0.000 & 0.000 & 0.055 & 0.036 & Loop (Small Bound)       \\
        Use Of Directly Dependent Variable                    & 0.010 & 0.009 & 0.000 & 0.064 & Assertion Satisfiability \\
        Use Of Indirectly Dependent (1) Variable              & 0.029 & 0.000 & 0.000 & 0.000 & Assertion Satisfiability \\
        Use Of Loop Condition Variable (Large Bound)          & 0.000 & 0.009 & 0.000 & 0.000 & Loop (Large Bound)       \\
        Use Of Loop Condition Variable (Small Bound)          & 0.010 & 0.000 & 0.000 & 0.000 & Loop (Small Bound)       \\
        Use Of Loop Condition Variable (Small Bound)          & 0.038 & 0.000 & 0.000 & 0.000 & Loop (Small Bound)       \\
        Use Of Recursion Condition Variable (Nondet Bound)    & 0.000 & 0.000 & 0.000 & 0.009 & Loop (Large Bound)       \\ \bottomrule
        \end{tabular}
}
\end{table}

\end{document}